\documentclass[iop]{emulateapj}

\usepackage{graphicx}
\usepackage{natbib}
\usepackage{txfonts}
\bibpunct{(}{)}{;}{a}{}{,} % to follow the A&A style
\usepackage{color} %only necessary for edits

\shorttitle{Structure of prominence legs}
\shortauthors{Levens et al.}

\begin{document}

   \title{Structure of prominence legs: 
Plasma and magnetic field}

   \author{P. J. Levens}
	\affil{SUPA School of Physics and Astronomy, University of Glasgow, Glasgow, G12 8QQ, UK}
	\affil{Observatoire de Paris, Meudon, 92195, France}
	\email{p.levens.1@research.gla.ac.uk}
 %\inst{1,2}
          %\inst{1}

	\author{B. Schmieder}
	\affil{Observatoire de Paris, Meudon, 92195, France}

 %\inst{2}
	\author{N. Labrosse}
	\affil{SUPA School of Physics and Astronomy, University of Glasgow, Glasgow, G12 8QQ, UK}
\and

 %\inst{1}
	  %\inst{1}
	\author{A. L\'{o}pez Ariste}
	\affil{Institut de Recherche en Astrophysique et Plan\'{e}tologie, Toulouse, France}

 %\inst{3}
          %\inst{1}\fnmsep\thanks{Just to show the usage
          %of the elements in the author field}

%   \institute{SUPA School of Physics and Astronomy, University of Glasgow,
%              Glasgow, G12 8QQ, UK\\
%              \email{p.levens.1@research.gla.ac.uk}
%         \and
%		Observatoire de Paris, Meudon, 92195, France
%	 \and
%		Institut de Recherche en Astrophysique et Plan\'{e}tologie, Toulouse, France
%         %    University of Alexandria, Department of Geography, ...\\
%         %    \email{c.ptolemy@hipparch.uheaven.space}
%         %    \thanks{The university of heaven temporarily does not
%         %            accept e-mails}
%             }

   \date{Received ...; accepted ...}

% \abstract{}{}{}{}{} 
% 5 {} token are mandatory
 
%  \abstract
\begin{abstract}

{We investigate the properties of a `solar tornado' observed on 15 July 2014, and aim to link the behaviour of the plasma to the internal magnetic field structure of the associated prominence. We made multi-wavelength observations with high spatial resolution and high cadence using \textit{SDO}/AIA, the IRIS spectrograph and the \textit{Hinode}/SOT instrument. Along with spectropolarimetry provided by the THEMIS telescope we have coverage of both optically thick emission lines and magnetic field information. AIA reveals that the two legs of the prominence are strongly absorbing structures which look like they are rotating, or oscillating in the plane of the sky. The two prominence legs, which are both very bright in \ion{Ca}{2} (SOT), are not visible in the IRIS \ion{Mg}{2} slit-jaw images. This is explained by the large optical thickness of the structures in \ion{Mg}{2} which leads to reversed profiles, and hence to lower integrated intensities at these locations than in the surroundings. Using lines formed at temperatures lower than 1 MK, we measure relatively low Doppler shifts on the order of $\pm$ 10 km s$^{-1}$ in the tornado-like structure. Between the two legs we see loops in \ion{Mg}{2}, with material flowing from one leg to the other, as well as counterstreaming. It is difficult to interpret our data as showing two rotating, vertical structures which are unrelated to the loops. This kind of `tornado' scenario does not fit with our observations. The magnetic field in the two legs of the prominence is found to be preferentially horizontal.
}

\end{abstract}

%The motivation of the study was to investigate the Dopplershifts of lines formed in
%cool plasma and see if they present the same duality of velocities as in coronal line Fe 195 A observed by EIS. in tornado like structure.

%We have observed a prominence in multiwavelength with high spatial resolution and high cadence due to the observations of the spectrograph IRIS and the Hinode/SOT.

%AIA and EIS reveal  that the two legs of the prominence has strong absorption structures which look to rotate with a velocity of 10 km/s as what is called "tornadoe" in recent papers.

%The results of the Mg II spectra analysis were surprising. The two legs,  very bright in Ca II (SOT),  are not  (or difficult) visible  in the Mg II SlJ images. This is explained by the optically thickness of the structures which lead to reverse profiles and the integrated intenity is much lower that the surrounding. With reverse profiles, the Dopplershifts are relatively low (max =0.5 km/s). Between the two legs we see very well the loops in Mg II with material flowing from one leg to the other and even with counterstreaming.
%The magnetic field in the two legs of the prominence are preferentially horizontal.
%It is difficult to interprete our data as two rotating  vertical structures not related to the loops. This kind of  scenario does not fit with these observations.

   \maketitle
%
%________________________________________________________________

\section{Introduction}
\label{sec:intro}

The word `tornado' has been used synonymously with prominences for many years \citep{Pettit32,Pettit43}, but the term has more recently been used again to describe specific phenomena in the solar atmosphere. \citet{Pike98} used the term to describe macrospicules seen by the \textit{Solar and Heliospheric Observatory} \citep[SOHO;][]{Domingo95}, however not in relation to a prominence. %I may have misunderstood this paper - But I didn't see any mention of prominence/filament with regards to these `tornadoes'
It was only with the launch of the \textit{Solar Dynamics Observatory} spacecraft and its high resolution imager, the Atmospheric Imaging Assembly \citep[\textit{SDO}, AIA;][]{Lemen12}, that authors began writing of `tornadoes' in relation to prominences again. \citet{Li12} and \citet{Su12} each used AIA images and techniques to demonstrate apparent rotation in prominence legs. \citet{Li12} studied a tornado-like prominence with relation to the surrounding cavity, and found swirling motions and apparent flows in the prominence structure{, and \citet{panesar13} investigated possible mechanisms for the observed motions and flows}. \citet{Su12} used time-distance diagrams, or `stack-plots', using coronal filters from AIA to measure and quantify sinusoidal oscillations in tornadoes, finding that the measured period related to a rotational velocity of 6--8 $\mathrm{km~s}^{-1}$.

The work of \citet{Wedemeyer12} was showing a swirling motion at different heights in the solar atmosphere, on disc, using both AIA and data from the Crisp Imaging Spectropolarimeter at the Swedish Solar Telescope \citep[CRISP, SST;][]{Scharmer03}. These `chromospheric swirls' they found to originate in the chromosphere, and extend part of the way into the corona in the AIA channels. These `magnetic tornadoes', however, do not appear to be related to any filamentary structure.

It is important here that we make a distinction between types of `tornado' as referenced in the literature. These `chromospheric swirls' have also been dubbed `magnetic tornadoes' \citep{Wedemeyer12,Wedemeyer13b}, but do not necessarily relate to prominences/filaments. Those that are prominence-tornadoes have been called `giant tornadoes' \citep{Wedemeyer13}, as well as more generally `solar tornadoes' \citep{Li12,Su12,panesar13,Su14,Levens15}, and it is these prominence-tornadoes that we are here referring to upon using this term. {However, we must make a further distinction here. The `solar tornadoes' which were studied in \citet{Su12,Su14} and \citet{Levens15} appear as absorbing features on the limb in AIA coronal images, whereas the one that was studied by \citet{Li12} and \citet{panesar13} appears bright in AIA coronal channels. This distinction is made in the work of \citet{Panasenco14}, where they argue that these observational features can be explained by 2D oscillation or counterstreaming in prominence spines and barbs, and plasma motion along a writhed field respectively. It is the first type that we study here.}

High resolution imaging can only build part of the picture of these structures. Spectroscopic techniques further our understanding of the nature of these solar tornadoes, and have been used by a number of authors. \citet{Wedemeyer13} presented the first Dopplergrams of such a structure using H-$\alpha$ data from CRISP/SST, where they found a split Doppler pattern with blueshifts of $20~\mathrm{km~s}^{-1}$ to $30~\mathrm{km~s}^{-1}$ on one side of the base of the tornado, and redshifts of $10~\mathrm{km~s}^{-1}$ to $30~\mathrm{km~s}^{-1}$ on the opposite side. More recently the Extreme-ultraviolet Imaging Spectrometer \citep[EIS;][]{Culhane07} on board the \textit{Hinode} \citep{kosugi_07} satellite has been used to study tornadoes, where \citet{Su14} found a similar split Doppler pattern to \citet{Wedemeyer13}, but this time in lines formed above 1.5 MK, with line-of-sight velocities of around 2--5 $\mathrm{km~s}^{-1}$. This temperature range was extended down to $\sim$1 MK by \citet{Levens15}. \citeauthor{Su14} found this antisymmetric Doppler pattern to be persistent for at least three hours of observation, which they and others \citep[e.g.][]{Orozco12} have interpreted as rotation of the tornado structure about a central axis.

Other explanations for these observations have been put forward \citep[see e.g.][]{Panasenco14}, with condensation along magnetic loops{, or oscillation and counterstreaming in prominence barbs}. \citeauthor{Wedemeyer13} identified giant tornadoes as the legs of prominences, which leads to the question: what is the magnetic structure of prominence legs? For active region filaments it appears that there is a common view that both ends of filaments are anchored in opposite sides of polarities along the polarity inversion line (PIL), with a spine making an angle of 25$^\circ$ with the PIL. For quiescent filaments, lying between low magnetic field regions, the situation is not so clear, and is still debated \citep{Mackay10}. The intermediate legs or feet would not be directly rooted in the polarities, but sustained in the dips of field lines of the magnetic rope of the filament, attracted towards the photosphere by parasitic polarities. In these regions the magnetic field is parallel to the photosphere, with the plasma in the dips \citep{Aulanier1998,Lopez2006,Dudik2008}. Analysis of both the magnetic structure of tornadoes and the physical conditions of the plasma is necessary for understanding the apparent rotation mechanism of the tornado{, and to discern whether or not it is really rotation that we are seeing}.

During an international campaign in 2014, utilising both ground-based and space-based instruments, we focused on the observation of prominences. We have here selected one prominence, observed on July 15, 2014, for analysis. This prominence appears to have similar observational characteristics to those of a tornado.

In Section \ref{sec:observations} we describe the instruments and observations: the Interface Region Imaging Spectrograph \citep[IRIS;][]{DePontieu2014}, the SOT instrument aboard \textit{Hinode}, as well as the polarimeter at the T\'{e}lescope H\'{e}liographique pour l'Etude du Magn\'{e}tisme et des Instabilit\'{e}s Solaires (THEMIS, French telescope in the Canary Islands). Also in Section \ref{sec:observations} we describe the prominence and tornado morphology as seen by \textit{SDO}/AIA. Section \ref{sec:parameters} discusses the physical parameters obtained from the analysis of the data from these various telescopes and what they tell us about the nature of the prominence. {Section \ref{sec:disc} contains discussion of our results within the context of previous work.} In Section \ref{sec:conc} we present our conclusions.

%__________________________________________________________________

\section{Observations}
\label{sec:observations}
\begin{figure}
\centering
\includegraphics[width=\hsize]{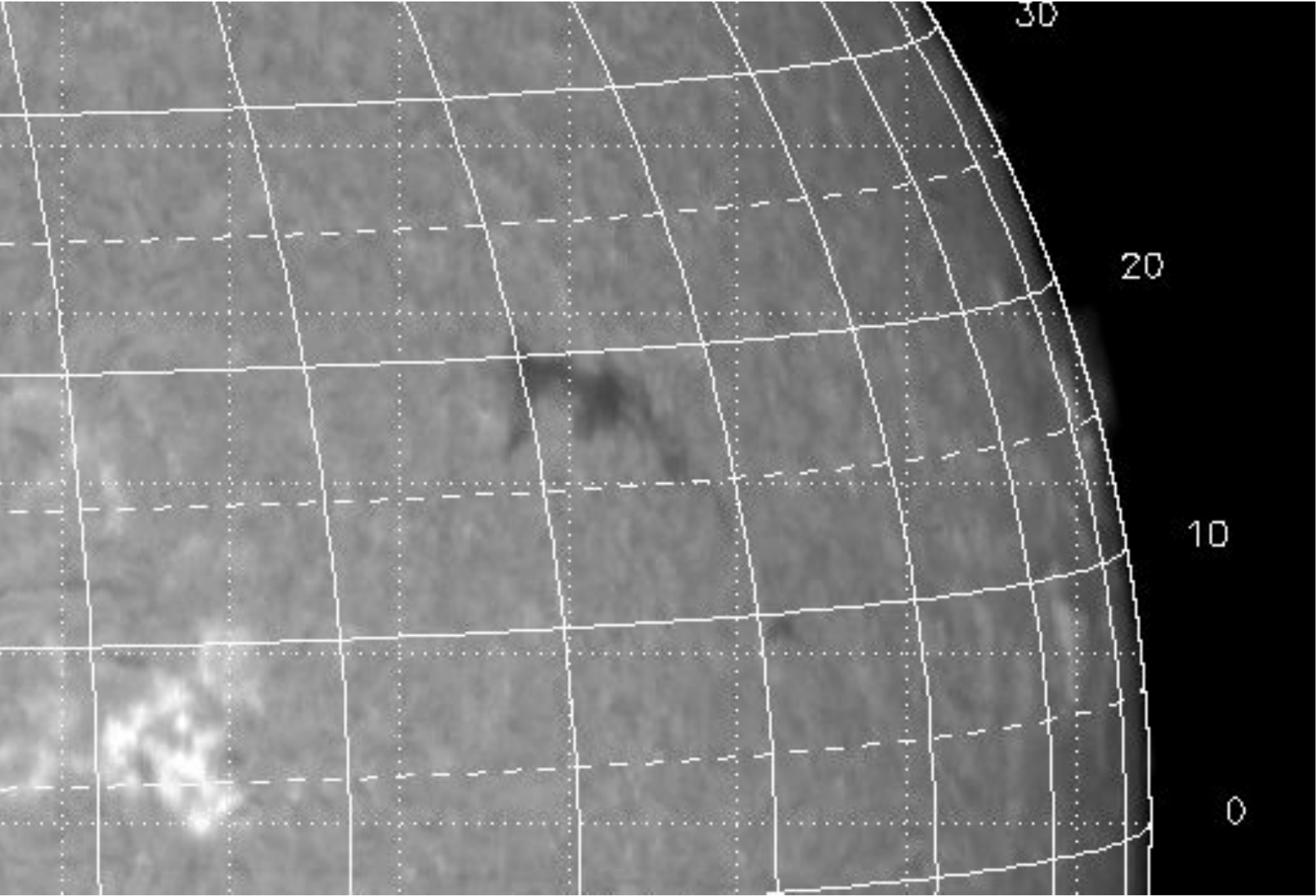}
\caption{Filament observed {at 13:37 UT} on July 11 2014, in H-$\alpha$ with the Meudon survey instrument.}
\label{Meudon_Ha_fil}
\end{figure}

\begin{figure}
   \centering
   \includegraphics[width=\hsize, trim=0 8cm 0 8cm, clip=true]{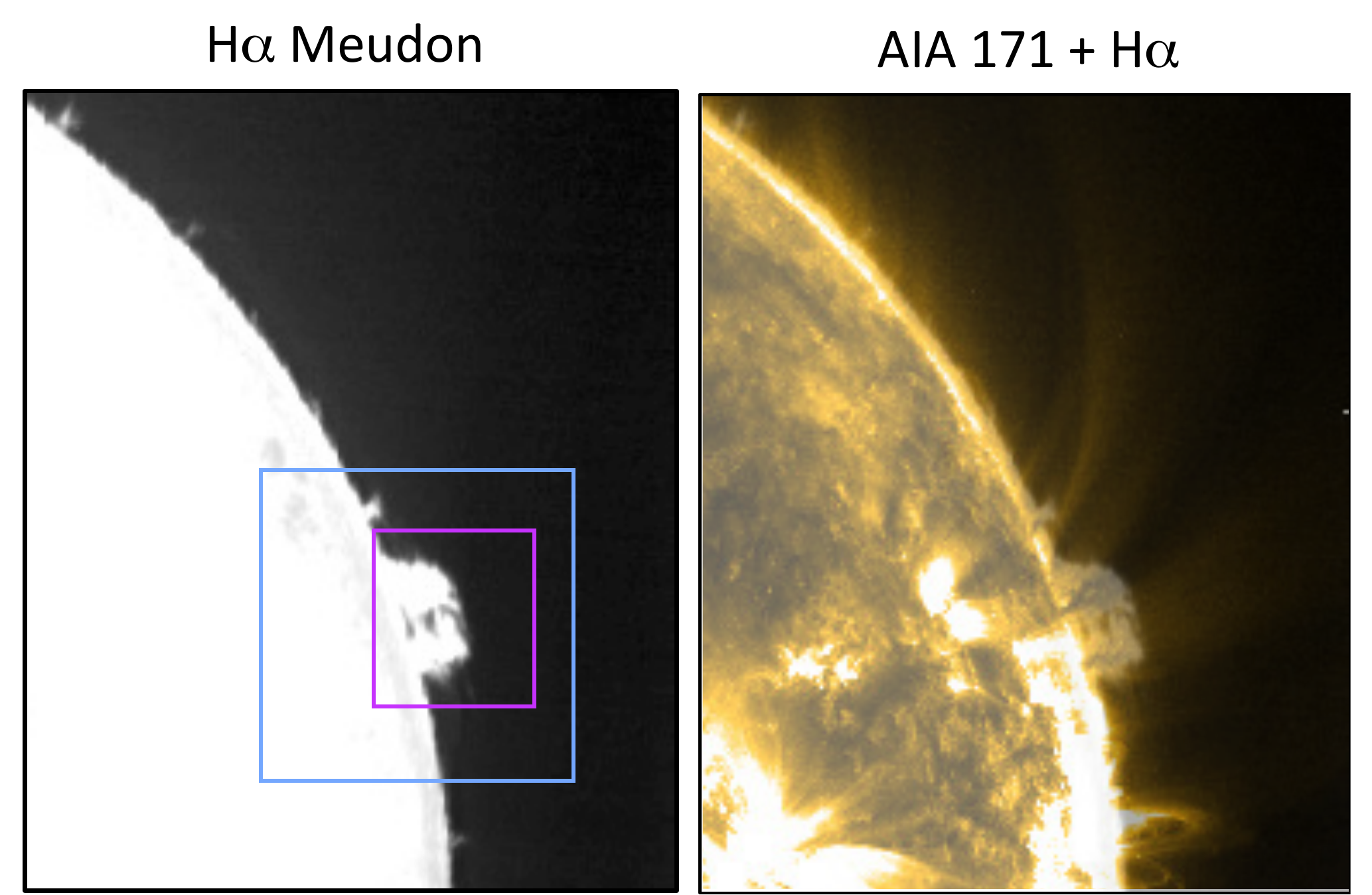}
      \caption{\textit{Left:} H-$\alpha$ prominence observed {at 09:40 UT} on July 15, 2014 (Meudon survey). {The purple and blue boxes show the fields of view in the AIA images of Figures \ref{AIA_171_304} and \ref{fig:iris_sot} respectively.} \textit{Right:} AIA 171 \AA\ overlaid with the H-$\alpha$ prominence.}
         \label{Meudon_171}
   \end{figure}

\begin{figure*}
\centering
\includegraphics[width=0.47\hsize,trim=0 5cm 0 5cm, clip=true]{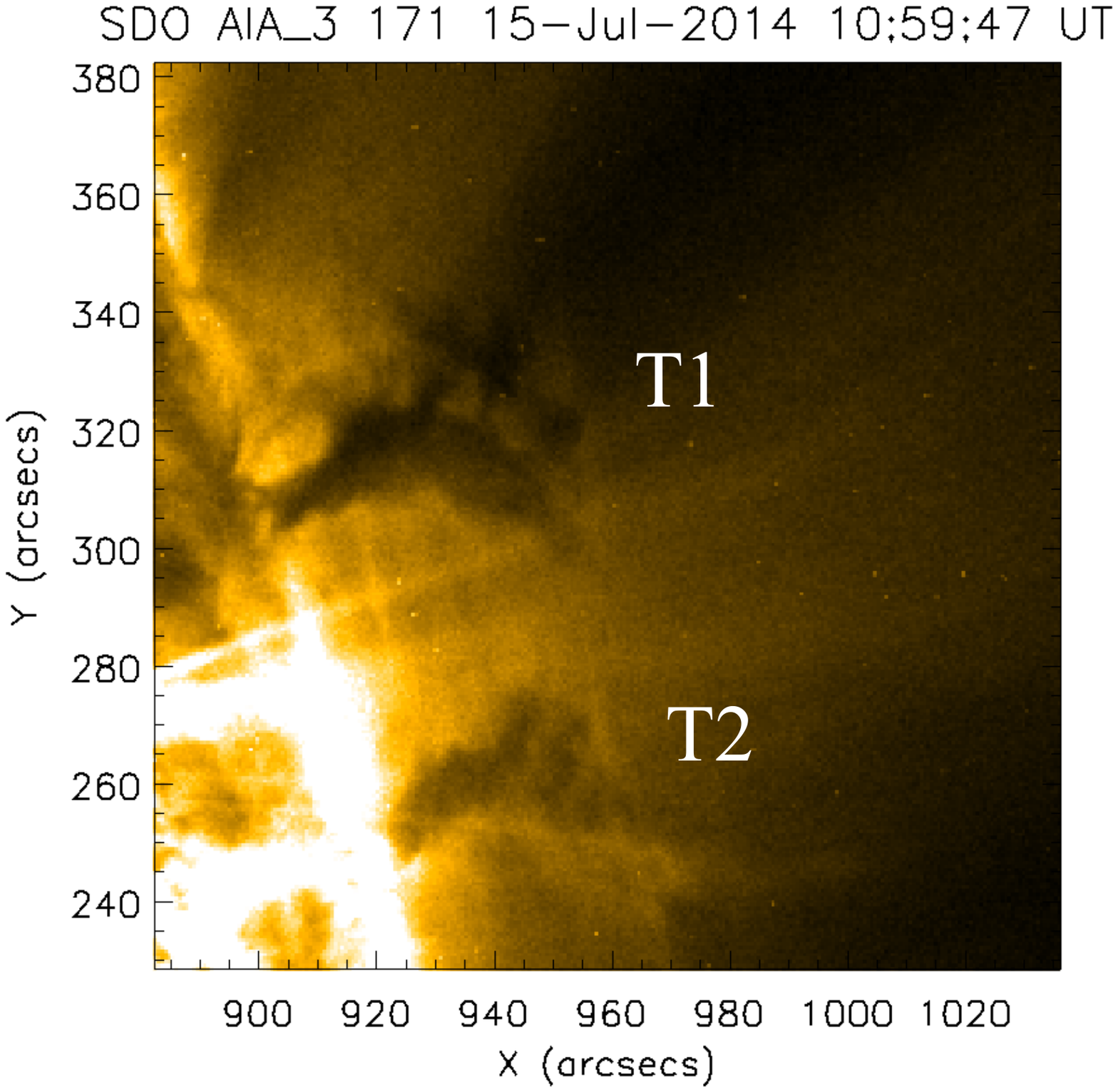}
\includegraphics[width=0.47\hsize,trim=0 5cm 0 5cm, clip=true]{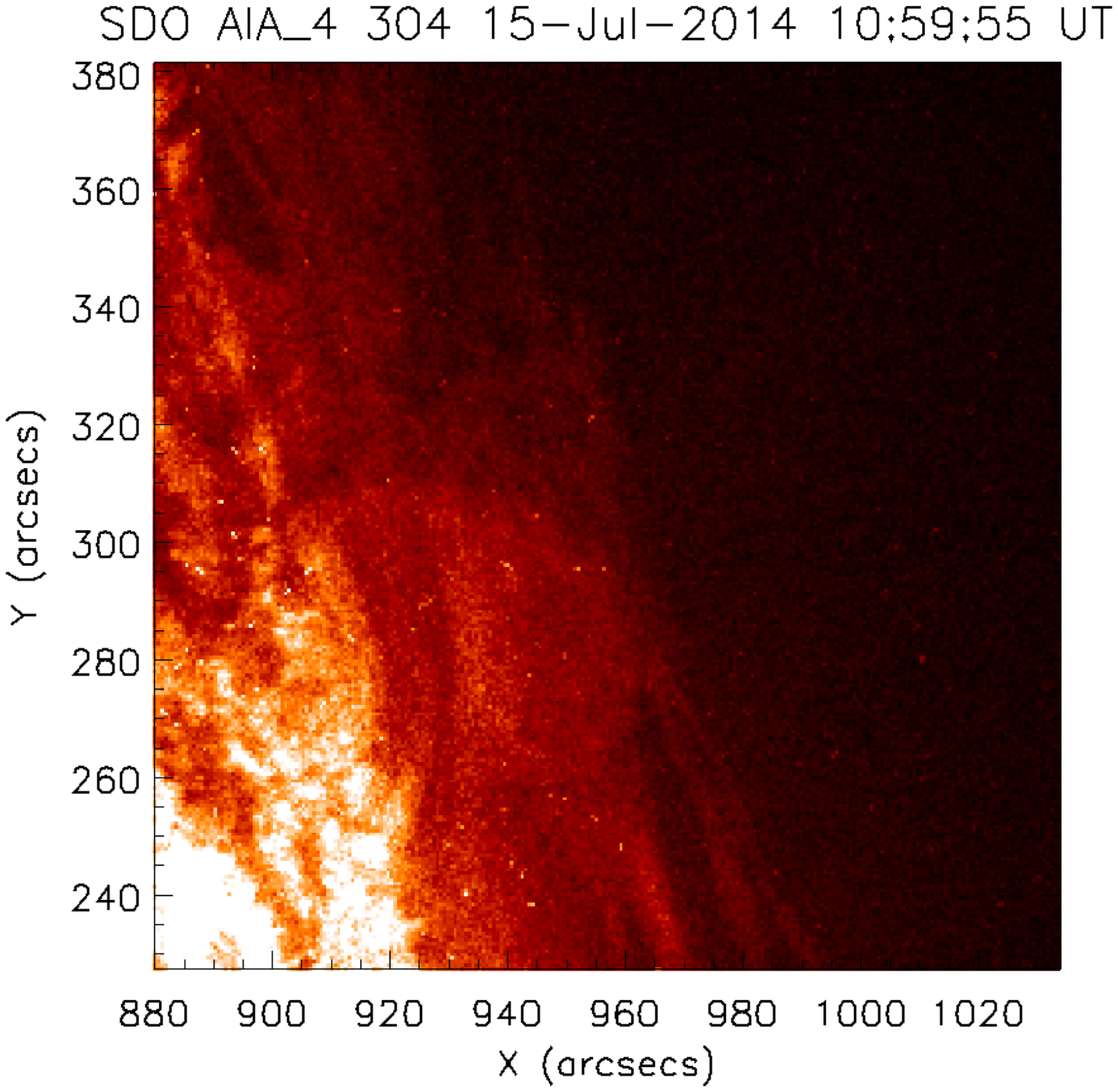}
\caption{The prominence observed on July 15, 2014 by AIA in 171 \AA\ (left) and 304 \AA\ (right). {Overlayed on the left panel are the identifiers used to denote the two columns here - T1 and T2. {Online movie shows the tornado-like nature of these two prominence legs in AIA 171 \AA.}}}
\label{AIA_171_304}
\end{figure*}

The target prominence for July 15 was tracked as filament across the disc for a number of days by ground-based observatories using H-$\alpha$ filters (e.g. from the Meudon survey, Figure \ref{Meudon_Ha_fil}, on July 11) before it became visible on the limb. Figure \ref{Meudon_171} (left panel) shows the prominence in H-$\alpha$ on July 15, clearly showing two column-like structures which are bright in emission. The right panel of Figure \ref{Meudon_171} shows an overlay of the AIA 171 \AA\ image from the same time. Images made using coronal filters of AIA (Figure \ref{AIA_171_304}, left panel {and online movie}) show the prominence as two dark columns of material, absorbing the background emission{, which display tornado-like behaviour when observed over time (see online movie accompanying Figure \ref{AIA_171_304})}. These dark columns from AIA align with the bright columns in H-$\alpha$ (Figure \ref{Meudon_171}, right panel).
  
%\begin{figure}
%\begin{center}
%\includegraphics[width=\hsize]{AIA_193_171_from_movie.pdf}
%\caption{Multi-waveband view of the prominence as seen by AIA. This image uses two coronal filters, 171 \AA\ and 193 \AA.}
%\label{fig:aia_movie_im}
%\end{center}
%\end{figure}

\begin{figure*}
\begin{center}
\includegraphics[width=0.6\textwidth,trim=0 6cm 0 6cm, clip=true]{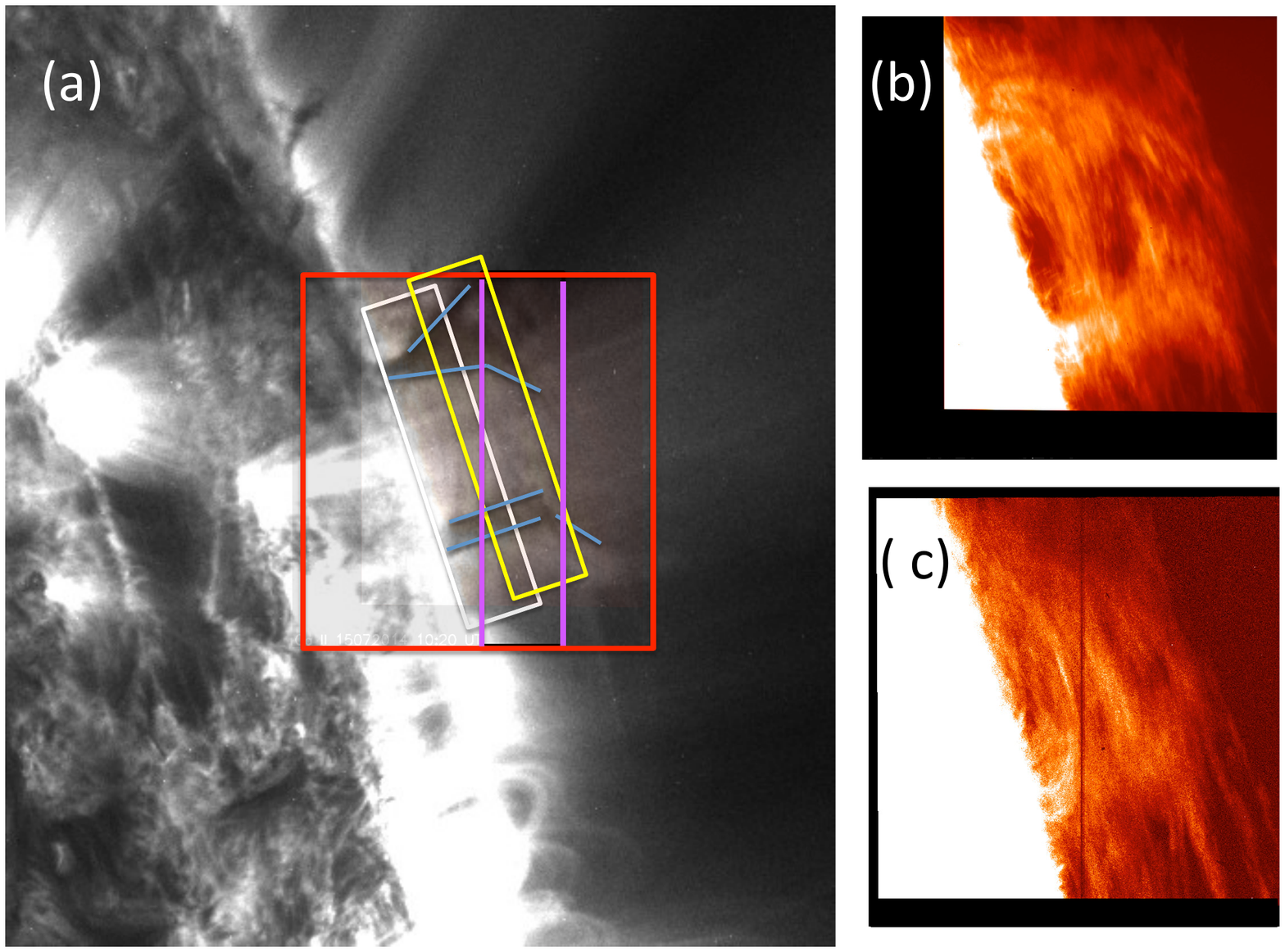}
\caption{{The coordinated prominence observation on 15 July 2014, here shown at 10:20 UT. (a) Two columns seen in absorption by AIA in the 171 \AA\ passband, with the field of view corresponding to that shown as a blue box in Figure \ref{Meudon_171}. Overlayed are the field of view of IRIS and SOT (red box) and the fields of view of the two THEMIS rasters from this day (white and yellow boxes). The magenta box outlines the region covered by the IRIS raster. The blue lines indicate the edges of the two absorption features. Co-aligned images of the prominence are shown in (b), \ion{Ca}{2} from \textit{Hinode}/SOT, and (c), in \ion{Mg}{2} from the IRIS slit-jaw using the 2796 \AA\ filter. The IRIS and SOT pointings were aligned by 2D cross correlation, using on-disc features. The images of SOT and IRIS (b and c respectively) are inserted into a black square, reflecting the red box of panel (a).}}
%\caption{The prominence  observed on 15 July 2014 at 10:20 UT  (a) seen in absorption in AIA 171 \AA, , overlaid with a square box (red) of IRIS and SOT field of view presented in b and c. The two rectangular  boxes (yellow and white)  are the two fields of view of THEMIS, the black box is the IRIS raster area for the spectra.  The blue lines are drawn at the edges of the two absorption features.  (b)  co-aligned image of the prominence  in Ca II  observed  by \textit{Hinode}/SOT  and  (c) in Mg II  line by  IRIS slit jaw. The SOT image was made using the \ion{Ca}{2} line, and the IRIS image is from the 2796 \AA\ filter, which is dominated by \ion{Mg}{2} emission. The IRIS and SOT pointings were aligned by 2D cross correlation using on-disc features. The two images of SOT and IRIS are inserted into a black square.}
%\caption{The prominence seen in absorption in AIA 171 \AA, observed on 15 July 2014 at 10:20 UT, overlaid with a square box (red) of IRIS and SOT field of view (a). The two rectangular  boxes (yellow and white)  are the two fields of view of THEMIS, the black box is the IRIS raster area for the spectra.  Also shown are co-aligned images of the prominence  by \textit{Hinode}/SOT (b) and the IRIS slit jaw (c). The SOT image was made using the \ion{Ca}{2} line, and the IRIS image is from the 2796 \AA\ filter, which is dominated by \ion{Mg}{2} emission. The IRIS and SOT pointings were aligned by 2D cross correlation using on-disc features.}
\label{fig:iris_sot}
\end{center}
\end{figure*}

The prominence appears differently when viewed in optically thick lines formed at chromospheric and transition region temperatures. Figure \ref{AIA_171_304}, right panel, shows the 304 \AA\ filter from AIA, which is dominated by emission from the \ion{He}{2} doublet at 303.78 \AA. In this waveband we see much more clearly the extent of the prominence above the limb, as well as the horizontal structure in the prominence body, which is also seen in \ion{Ca}{2} from \textit{Hinode}/SOT (Figure \ref{fig:iris_sot} (b)) and \ion{Mg}{2} from IRIS (Figure \ref{fig:iris_sot} (c)).

\subsection{Instruments}
\label{ssec:inst}

\subsubsection{IRIS}
\label{sssec:iris}
IRIS performed a 16-step coarse raster observation from 08:00 to 11:00UT on July 15, 2014. The pointing of the telescope was (940\arcsec, 284\arcsec), with spatial pixel size of 0.167\arcsec. The raster cadence of the spectral observation in
both the near ultraviolet (NUV, 2783 to 2834 \AA) and far ultraviolet (FUV, 1332-1348 \AA\ and 1390-1406 \AA)
wavelength bands was 86 seconds. Exposure time was 5.4 seconds per slit position. Slit-jaw images (SJI) in the broadband filters 
(2796 \AA\ and 1330 \AA) were taken with a cadence of 11 seconds. The FOV was 30\arcsec$\times$119\arcsec\ for the raster and 119\arcsec$\times$119\arcsec\ for the SJI. Calibrated level 2 data is used for this analysis, with dark current subtraction, flat field correction, and geometrical correction having all been taken into account \citep{DePontieu2014}.

We mainly used the \ion{Mg}{2} k 2796.35~\AA\ and \ion{Mg}{2} h 2803.5~\AA\ lines, along with the slit-jaw images in the 1330 \AA\ and 2796~\AA\ filters for this study. The \ion{Mg}{2} h and k lines are formed at chromospheric plasma temperatures ($\sim$10$^4$ K). The SJI 2796 \AA\ filter samples emission mainly from the \ion{Mg}{2} k line, while emission in the 1330~\AA\ slit jaw is an integration of the FUV emission from within a range of about 40~\AA, including the total emission of two \ion{C}{2} lines and UV continuum. 
%formed in the lower chromosphere and the Si IV 1402 and1393~\AA\ lines formed in the prominence transition region (PCTR). 
The co-alignment between the optical channels is achieved by comparing the positions of horizontal fiducial lines.

\subsubsection{\textit{Hinode}/SOT}
\label{sssec:sot}
The \textit{Hinode} Solar Optical Telescope \citep[SOT;][]{tsuneta_08,suematsu_08} consists of a 50 cm diffraction-limited Gregorian telescope and a Focal Plane Package including the narrowband filtergraph (NFI), broadband filtergraph (BFI), the Stokes Spectro-Polarimeter, and Correlation Tracker (CT). For this study, images were taken with a 30 second cadence in the \ion{Ca}{2} H line at 3968.5~\AA\ using the BFI. The \ion{Ca}{2} images have a pixel size of 0.109\arcsec, with a field of view of 112\arcsec$\times$112\arcsec.

\subsubsection{THEMIS spectropolarimetry}
\label{ssec:themis}

The slit of the THEMIS MulTi Raies (MTR) spectrograph \citep{LopezAriste00} was orientated parallel to the limb for these observations. The observations consist of two successive rasters with 30 positions separated by 2\arcsec. The first raster started at 14:41 UT, and the second began at 16:55 UT. The acquisition of the full raster takes more than one and a half hours. For an unknown reason the rasters each contain only 11 slit positions, covering a 22\arcsec\ region. The slit is 120\arcsec\ long, and the pixel size is 1\arcsec. Due to the grid mode of the polarimeter, the images are obtained by taking exposures at two successive displacements of the grid along the slit in order to cover the full spectra of the 4 Stokes parameters in the \ion{He}{1} D$_3$ line (I, Q, U and V). The nature of this observing mode creates dark vertical bars in the images, which indicate the boundaries between grid placements.

%The raw data of the THEMIS/MTR mode was reduced using the DeepStokes procedure \citep{LAARMSDG09} and the Stokes profiles were fed to an inversion code based on Principal Component Analysis \citep{LAC02,Casini03}. The observed profiles were compared against those in a database containing 90000 profiles, generated with known models of the polarization profiles of the \ion{He}{1} D$_3$ line and taking into account the Hanle and Zeeman effects \citep{LAC02}. The details of the MTR data reduction can be found in \citet{Schmieder2013,Schmieder14}. The most similar profile in the database is kept as the solution, and the parameters of the model used in the computation are kept as the inferred vector magnetic field, height above the photosphere and scattering angle. Error bars on these parameters are determined by performing statistics on all other models which are similar to the observed profile. %, but not as similar as that which is selected as the solution.

{The raw data of the THEMIS/MTR mode was reduced using the DeepStokes procedure \citep{Lopez09}. The details of this data reduction can be found in \citet{Schmieder2013,Schmieder14}. The resulting Stokes profiles of each pixel were fed to an inversion code based on Principal Component Analysis \citep{Lopez02,Casini03}. In this inversion technique the observed profiles are compared to those in a database containing 90000 profiles, which were generated with known models of the polarization profiles of the \ion{He}{1} D$_3$ line, taking into account the Hanle and Zeeman effects \citep{Lopez02}. The optical thickness of the \ion{He}{1} D$_3$ line in prominences is small \citep{Labrosse01,Labrosse04}, and this justifies the single scattering approximation used in the computation of the profiles. It is also worth remembering that polarization is only dependent on population imbalances inside the 5-level atomic model used for the calculations but not on the total atomic population of this model. This so-called non-LTE problem of the second kind \citep{Landideglinnocenti04} is solved, but in the absence of important optical thickness there is no need to address other non-LTE effects.

Returning to the inversion process, the most similar profile in the database is kept as the solution, and the parameters of the model used in the computation are kept as the inferred vector magnetic field, height above the photosphere and scattering angle. As described by \citet{Lopez02}, a set of the nearest models, including the nearest one given as solution, is used to determine, for each inverted point, error bars for each one of the parameters of the model. These error bars are explicitly computed as the standard deviation of each one of the parameters of the model in that set of models near to the solutions. Four factors contribute to these error bars: the noise in the observed profiles, the finite size of the database, the presence of inherent ambiguities (as e.g. the 180 degrees ambiguity in the case of the azimuth) and the ability of our model to reproduce the observations. Ambiguities are well-known and their impact has been studied by \citet{Casini05}. \citet{Lopez02} found the typical error bars coming from the two first issues, noise and finite size of the database. Of particular interest for the present work is the error bar they found for the inclination of the magnetic field, 10 degrees. Any inversion error in the retrieved inclination larger than 10 degrees can therefore only be attributed to the fourth cause: the insufficiency of the model to explain the observations. \citet{Schmieder15IAU} and \citet{Lopez15IAU} already used this fact to look for complex magnetic topologies and we shall return to this point in Section \ref{ssec:mag_field} below.}

%Figure \ref{themis2} presents the maps obtained after the inversion of the Stokes parameters recorded in the \ion{He}{I} D$_3$ line: (a) Intensity, (b) magnetic field strength, (c) inclination, and (d) azimuth for the scan represented by the yellow box in Figure \ref{fig:iris_sot}. The origin angle of inclination is the local vertical, and the origin of the azimuth is the line of sight (LOS) in a plane containing the LOS and the local vertical. We see that the brightest parts of the prominence have a mean inclination of 90$^\circ$ which means that the magnetic field in these parts is horizontal. However there is a large dispersion of the values ($\pm$30$^\circ$) from one pixel to the next in the lateral part of the prominence.

%______________________________________________________________

\subsection{Morphology of the tornado}
\label{ssec:morphology}

\subsubsection{AIA, SOT and IRIS}
\label{sssec:aia}
The \textit{SDO}/AIA images for the event on July 15 2014 show different structures in different wavebands, seen in Figure \ref{AIA_171_304}. It is in the AIA coronal filters, such as the left panel of Figure \ref{AIA_171_304}, that we see the apparent turning of these structures, leading to the impression of a ``tornado'' {(see online movie associated with Figure \ref{AIA_171_304})}. In the 171 \AA\ images (Fig. \ref{AIA_171_304}, left panel) we see the tornadoes as `silhouettes' against the hot ($\sim$ 1 MK) background coronal emission. The orientation of these two columns on the limb is not as simple as it seems from this AIA image. The prominence is not orientated north-south on the limb, rather it appears to have more of an east-west orientation (as can be seen in Figure \ref{Meudon_Ha_fil}). The northern column (herein known as tornado 1, T1{, as labelled in Figure \ref{AIA_171_304}, left panel}) is located closer to the observer than the southern one (herein tornado 2, T2{, as shown in the lower part of Figure \ref{AIA_171_304}, left panel}). This can be seen in AIA movies of the days leading up to this observation, in which T2 clearly crosses the limb first. % Do I need to put in a link to online movies or something?
We also note that there are loops of hot material in front of tornado 2, slightly obscuring it in these images. This is especially clear in other AIA coronal images (e.g. 193 \AA\ and 211 \AA), where T2 is completely obscured by the hot foreground emission. % Put these images in? Or not bother saying this at all?

The northern of the two columns consists of a vertical pillar of 10\arcsec\ wide and 15\arcsec\ tall, extending towards 50\arcsec\ high with different branches which are moving around and changing shape. These branches occupy an inverted cone which is 40\arcsec\ wide at the top. The southern column is a wide pillar, perpendicular to the limb (10\arcsec\ wide, 20\arcsec\ tall).

In the AIA 304 \AA\ image (Fig. \ref{AIA_171_304}, right panel) we see a different structure. The prominence is much more extended than those absorbing features seen in 171 \AA\ \citep{Labrosse11}, and appears similar in 304 \AA\ emission to both IRIS and \textit{Hinode}/SOT images. We also note that there is a darker structure in the 304 \AA\ image at the location of tornado 1 - additional absorption that is not visible in the IRIS \ion{Mg}{2} images. We note that this extra absorption could be due to other, hotter lines in this AIA passband. Although this waveband is dominated by \ion{He}{2} emission there are also a number of lines formed at coronal temperatures in the window, most notably \ion{Si}{11} at 303.3 \AA, formed at $\log{\mathrm{T}} = 6.2$ \citep{Dere97,Landi12}.

Figure \ref{fig:iris_sot}(b) shows the prominence as seen by SOT in \ion{Ca}{2}. In this image not only can we identify the two columns which are visible in AIA 171 \AA, but we also see some loop-like structure, similar to that seen in AIA 304 \AA\ (Fig. \ref{AIA_171_304}, right panel) and the IRIS {\ion{Mg}{2}} SJI (Fig. \ref{fig:iris_sot}(c)).

In Figure \ref{fig:iris_sot}(c) we show the IRIS slit-jaw image of the prominence using the 2796 \AA\ filter. The red box in Figure \ref{fig:iris_sot} (a) indicates the placement of the IRIS SJI with respect to the background AIA 171 \AA\ image, and the magenta box shows the rastering region that was covered during the study. The two tornadoes are not clearly visible in this waveband. This will be discussed in Section \ref{sssec:gas_pressure}, after inspecting the spectra - both \ion{Mg}{2} and \ion{Ca}{2} lines have a large optical thickness compared to H-$\alpha$.

\subsubsection{Time-distance analysis}
\label{sssec:time_distance}

\begin{figure}
\begin{center}
\includegraphics[width=0.9\hsize, trim=4.5cm 7.5cm 0 5cm, clip=true]{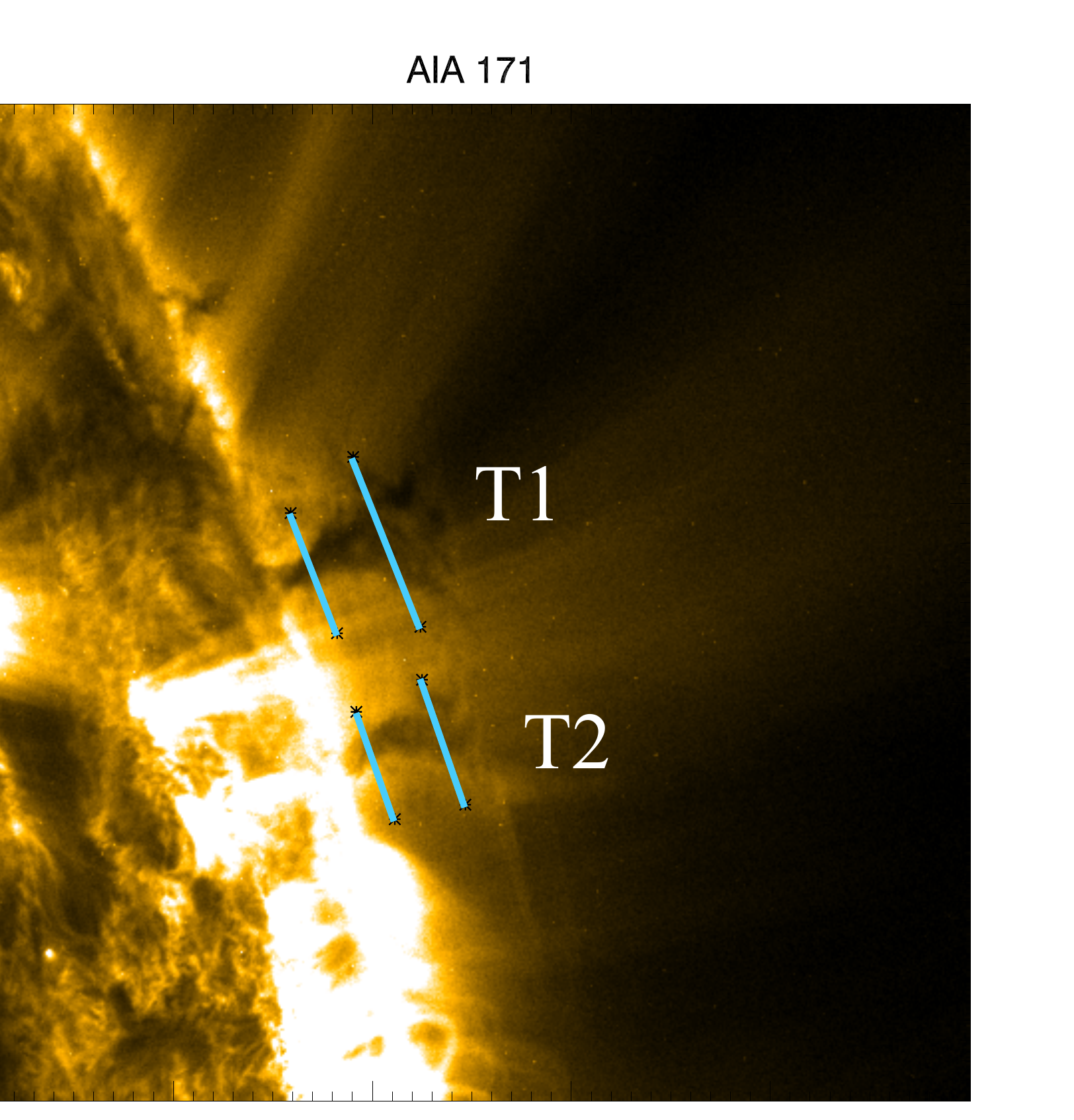}
\\
\includegraphics[width=0.7\hsize, trim=1.8cm 0 0 0, clip=true]{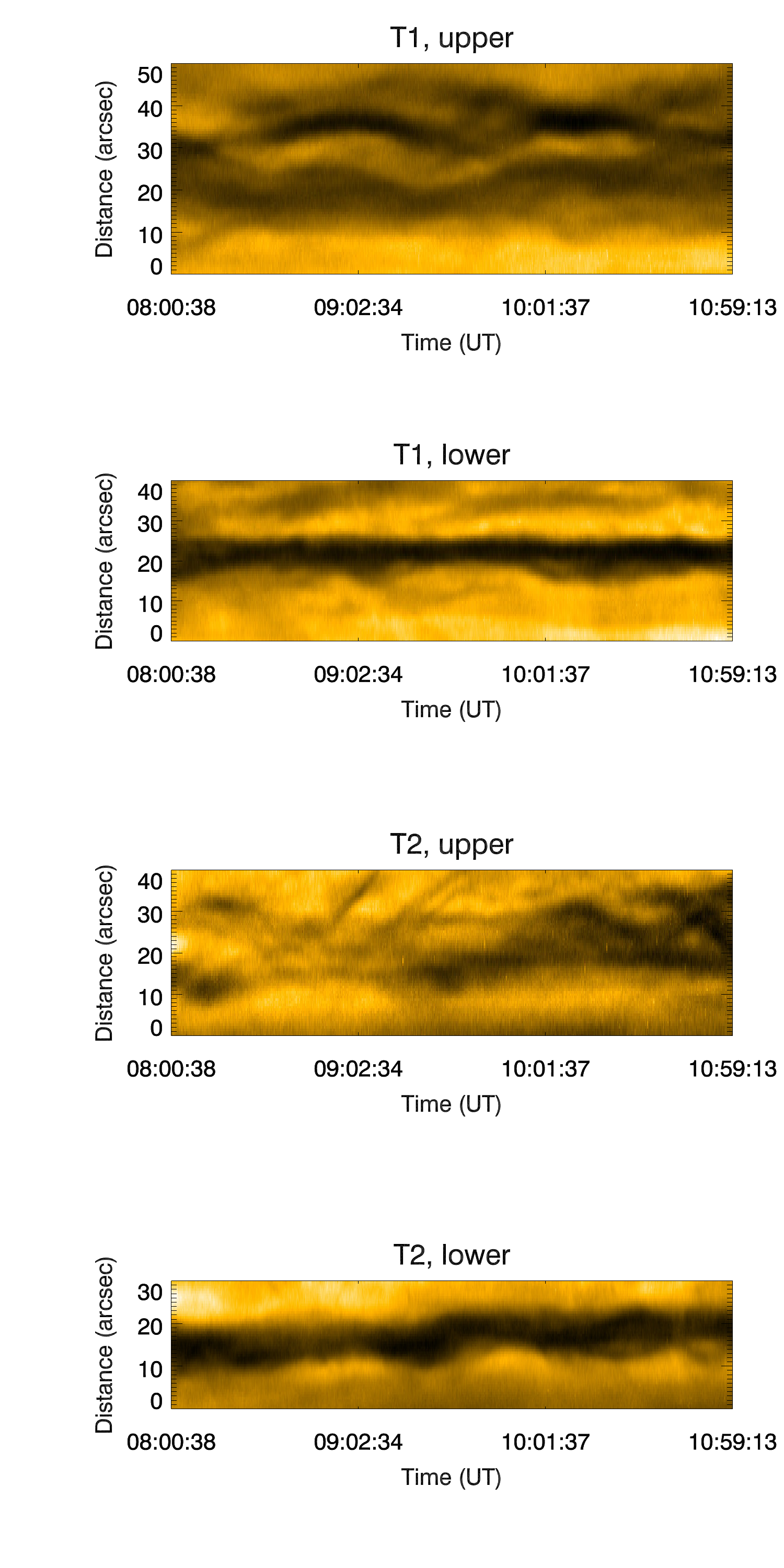}
\caption{Time-slice diagrams for the two tornadoes, T1 in the north and T2 in the south, using the AIA 171 \AA\ waveband. {The top image was taken at 11:00 UT, corresponding to the end time of the IRIS observation.} Here we have taken two parallel cuts through each column: a lower cut near the limb and an upper cut which is nearer the top of each tornado, {and these are shown in blue in the top panel}.}
\label{fig:AIA_tornado_cuts}
\end{center}
\end{figure}

As was mentioned in Section \ref{sec:intro}, \citet{Su12} used time-distance plots to measure sinusoidal oscillations in tornado-like prominences. In that paper they found a period that would relate to a rotational velocity of 6--8 km s$^{-1}$. We apply a similar method here to the two tornadoes in the AIA 171 \AA\ image, the results of which are presented in Figure \ref{fig:AIA_tornado_cuts}. Here we have taken two cuts through each of the columns, parallel to the limb, one at the base, and one near the top of the absorbing feature.

We selected different cuts across the two columns, parallel to the limb, to perform the time-distance analysis. The lower set of cuts are approximately 10\arcsec\ above the limb, with the upper cuts at 35\arcsec\ from the limb.
The cuts through the lower part of each of the columns do not appear to display any oscillatory motion in the plane of the sky, or at least there is not enough contrast in the AIA images to measure. The upper parts of both columns, however, display a much clearer sinusoidal pattern. Taking a rough estimate at the period {($\sim$1.25 hours)} and amplitude {($\sim$10\arcsec)} of these oscillations, we recover a velocity on the order of 10 km s$^{-1}$, which is consistent with the results of \citet{Su12}{, if we assume that the tornado is rotating} 

{However, as has been pointed out by \citet{Panasenco14}, time-distance analysis like this does not prove that the observed structure is rotating. These observational signatures could also be caused by oscillations of plasma in the magnetic field, projected onto the plane of the sky. We cannot rule out this explanation with this data set, as will be discussed in the following analysis of THEMIS and IRIS data.}

%______________________________________________________________

\section{Physical parameters}
\label{sec:parameters}

\subsection{Magnetic field vector}
\label{ssec:mag_field}

\begin{figure}
%\centerline{\includegraphics[width=\hsize,trim=1cm 0 2cm 0,clip=true]{July15_t038_map}}
\centerline{\includegraphics[width=\hsize,trim=1cm 7cm 2cm 8cm,clip=true]{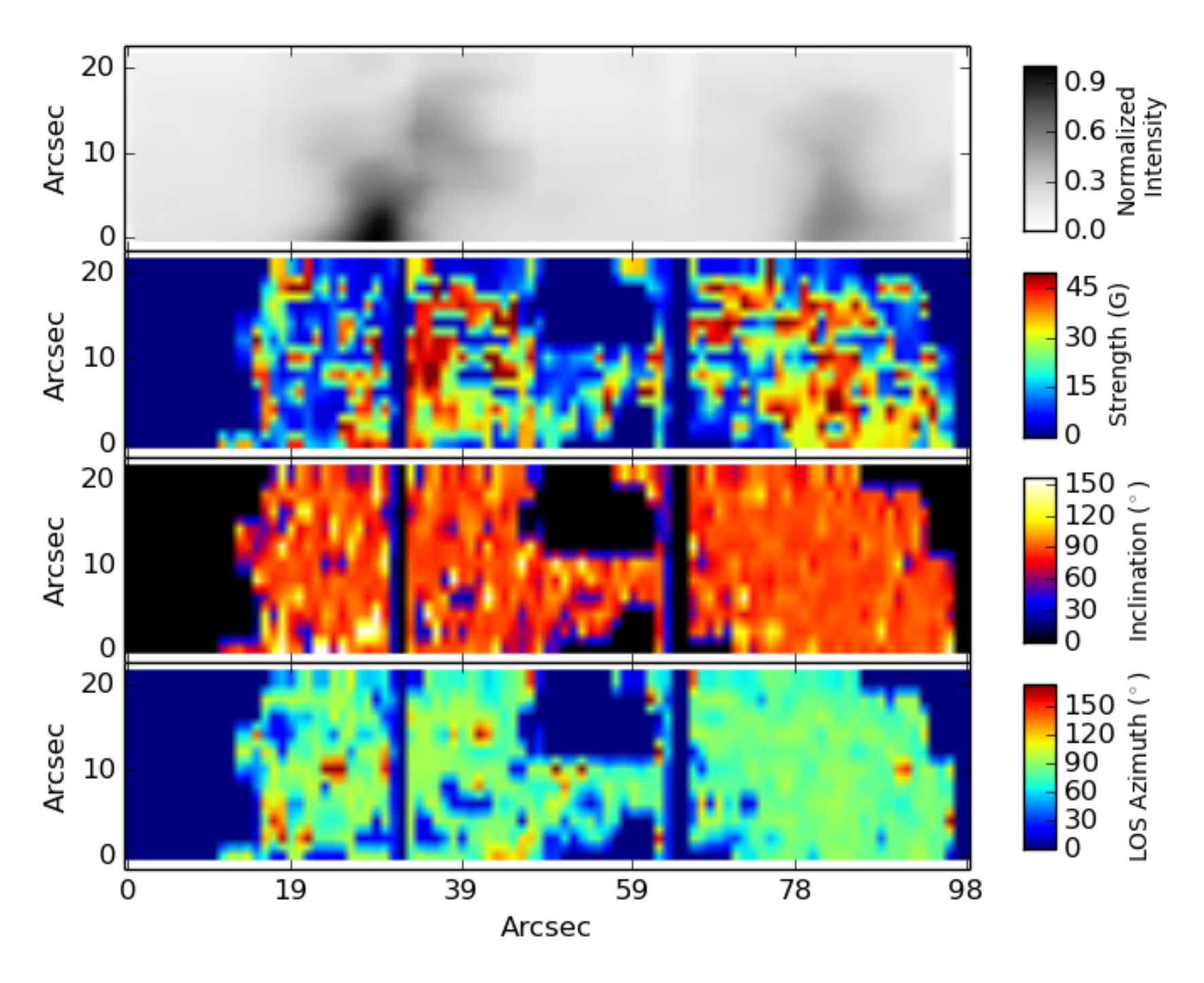}}
%}
 \caption{THEMIS observation of the prominence on July 15, 2014 between 16:55 UT and 18:10 UT in \ion{He}{1} D$_3$: ({\it from top to bottom})  intensity, magnetic field strength, inclination and azimuth.}
 % The two  lines (dot-dashed and dashed) in the intensity image represent the location of the two  cuts used  in Figure \ref{themis3}. The origin of the cut in the brightest region is x=40, y=0, the origin of the IRIS slit is x-80, y=0.
 \label{themis2}
\end{figure}

The He I D$_3$ 5876 \AA\ line appears as  a doublet. The two spectroscopic components have a different sensitivity to the Hanle effect and are far from being saturated. This helps with constraining the two ambiguities that need to be resolved: the 180 degree ambiguity of the Zeeman effect and the 90 degree ambiguity of the Hanle effect. The first raster overlays the bottom 22\arcsec\ of the two columns, while the second raster is aimed at the top, with a slight overlap with the first raster - see Figure \ref{fig:iris_sot} (a), where the upper and lower THEMIS raster positions are represented by white and yellow boxes respectively.

Figure \ref{themis2} presents the maps obtained after the inversion of the Stokes parameters recorded in the \ion{He}{1} D$_3$ line: ({\textit{from top to bottom}}) Intensity, magnetic field strength, inclination, and azimuth for the scan represented by the yellow box in Figure \ref{fig:iris_sot} (observed between 16:55 UT and 18:10 UT). The origin angle of inclination is the local vertical, and the origin of the azimuth is the line of sight (LOS) in a plane containing the LOS and the local vertical. We see that the brightest parts of the prominence have a mean inclination of 90$^\circ$ which means that the magnetic field in these parts is horizontal. However there is a large dispersion of the values ($\pm$30$^\circ$) from one pixel to the next in the lateral part of the prominence. {Also in these bright parts of the prominence we find the strongest magnetic fields, ranging from around 20 G up to 50 G in some regions. We recover l.o.s. azimuth values of between 70$^\circ$ and 100$^\circ$.} {We note there is some structure between the columns, visible in the bottom three panels of Figure \ref{themis2}, seemingly joining the columns. In the intensity map, this structure is very dim, and is not visible without over-saturating the main prominence pillars. It is not certain that this faint structure is part of the prominence, or if it is plasma sitting in the foreground or background of the prominence that we are studying. It does not, however, appear to play any role in the inversions that we are interested in here, which is those from the columns themselves.}

%%%%Figure of the first raster should be added.

\begin{figure}
\begin{center}
\includegraphics[width=\hsize]{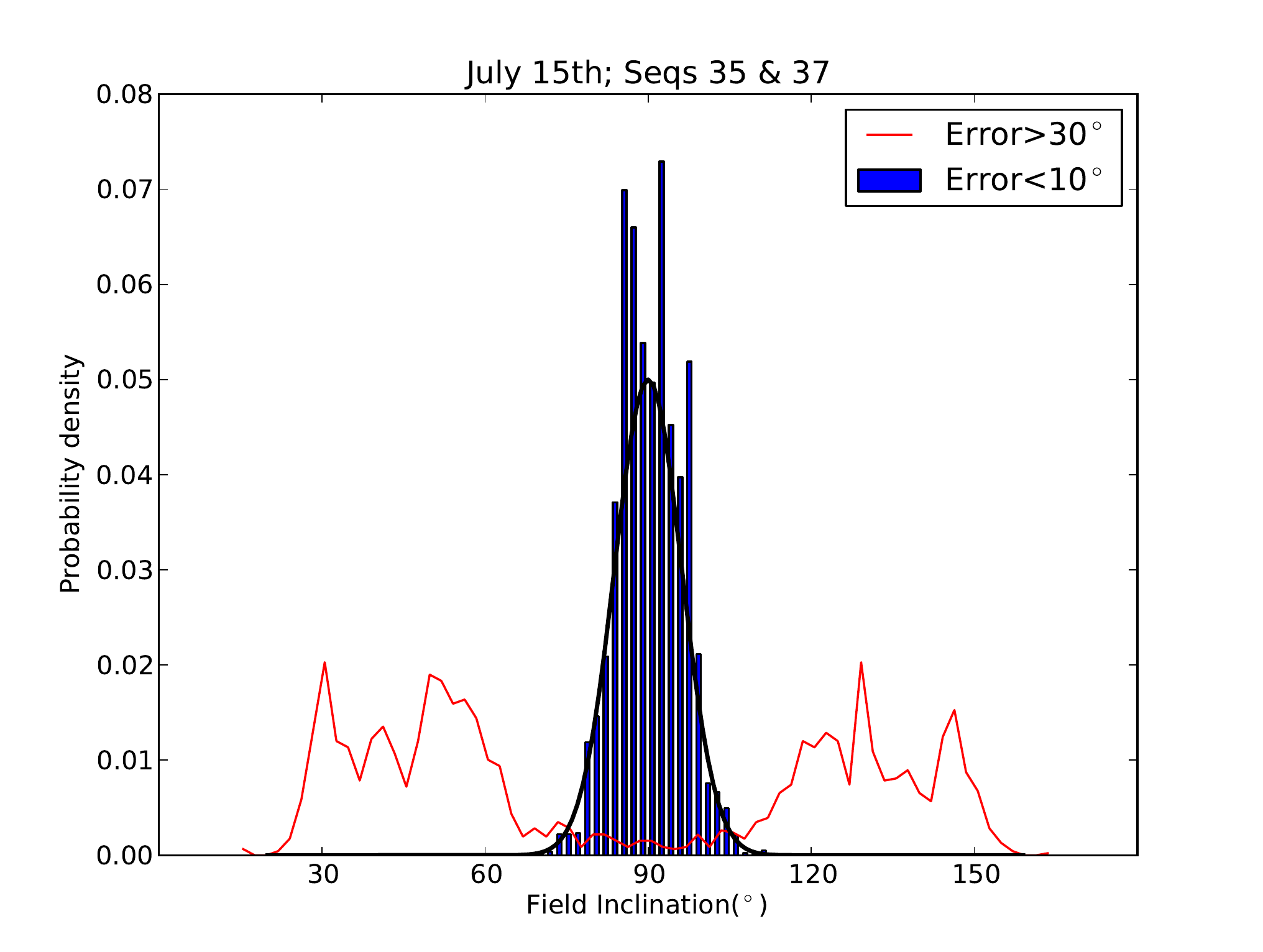}
\caption{Histogram of magnetic field inclination for points in the prominence. The blue histogram indicates points whose error is $<$ 10$^\circ$, with the black line is a Gaussian with FWHM = 10$^\circ$. The red line indicates the inclination of points where the error is $>$ 30$^\circ$.}
\label{fig:histo}
\end{center}
\end{figure}

%We have computed the histogram of the inclination for all points in the tornado, shown in Figure \ref{fig:histo}. The main peak of the distribution is centered on 90$^\circ$, indicating a horizontal field with respect to the local limb. Those results have small error bars and we are confident that the magnetic field of the prominence is correctly measured. There are also two secondary peaks in Figure \ref{fig:histo}, whichs are wide - from 30$^\circ$ to 60$^\circ$ and from 120$^\circ$ to 150$^\circ$ - with large error bars ($>$ 30$^\circ$). The inclinations of 60$^\circ$ and 120$^\circ$ have already been detected in prominences \citep{Schmieder14,Lopez15IAU}%{\bf (NEED TO ADD REFERENCES TO SCHMIEDER AND LOPEZ IAU PROCEEDINGS - Where is this available? )}
%. These inclinations have been explained by the superposition of two magnetic fields, one horizontal and the other turbulent. The inclinations of 30$^\circ$ and 150$^\circ$, on the other hand, have also been observed in other tornadoes \citep{Schmieder15IAU}. We concluded from this that the magnetic field model of one field per pixel used by the PCA inversion code is not valid. We can definitely exclude a vertical field as a solution, as this would have been correctly inverted. Several possibilities that mix a number of magnetic components are yet to be explored.% as in the case of the prominence have to be  explored.

{We have computed the histogram of the inclination for all points in the tornado, shown in Figure \ref{fig:histo}. The main peak of the distribution is centered on 90$^\circ$, indicating a horizontal field with respect to the local limb. Those results have small error bars of 10 degrees (the black line is a Gaussian curve with standard deviation of 10 degrees that correctly reproduces the data distribution). As described above, this error bar is consistent with the expected errors from noise and the finite size of the database. We are confident that the magnetic field of the prominence is correctly measured in those cases. There are also two secondary peaks in Figure \ref{fig:histo}, which are wide - from 30$^\circ$ to 60$^\circ$ and from 120$^\circ$ to 150$^\circ$ - with large error bars ($>$ 30$^\circ$). These large error bars cannot be attributed either to noise, the finite size of the database or ambiguities. The inclinations of 60$^\circ$ and 120$^\circ$ have already been detected in prominences \citep{Schmieder15IAU,Lopez15IAU}. These inclinations have been explained by the superposition of two magnetic fields: one horizontal and the other turbulent. This magnetic topology is not considered in the database used for inversion that includes just one magnetic field vector which is constant over the pixel. The inversion code tries to fit with a spread of models resulting in such error bars in these cases. The inclinations of 30$^\circ$ and 150$^\circ$, on the other hand, have also been observed in prominences \citep{Schmieder14}. We concluded from this that the magnetic field model of one field per pixel used by the PCA inversion code is not valid. We can definitely exclude a vertical field as a solution, as this would have been correctly inverted. Several possibilities that mix a number of magnetic components are yet to be explored.}

%______________________________________________________________

%\subsection{Spectral analysis of EIS}
%\label{ssec:eis_analysis}

%______________________________________________________________

\subsection{Spectral analysis of IRIS}
\label{ssec:iris_analysis}
\begin{figure}
\begin{center}
\includegraphics[width=\hsize]{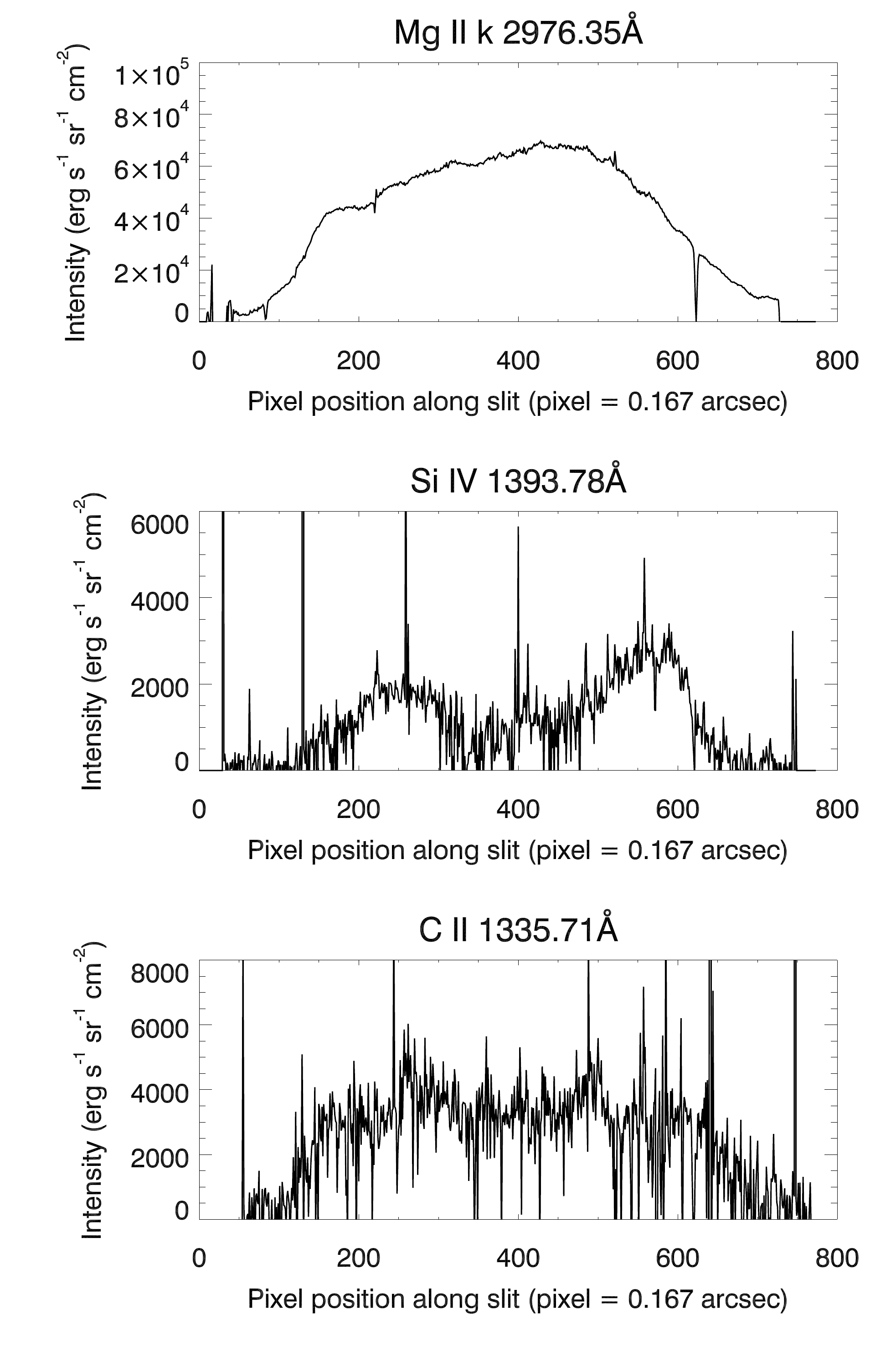}
\caption{Intensity as a function of slit position for three IRIS spectral lines from the raster - \ion{Mg}{2} k 2796.35 \AA\ (top), \ion{Si}{4} 1393.78 \AA\ (middle) and \ion{C}{2} 1335.71 \AA\ (bottom). These have all been averaged over an hour's worth of data in order to combat the poor signal-to-noise ratio of the \ion{Si}{4} and \ion{C}{2} lines. Also note the differences in scale on the y-axes. These are all taken from the third IRIS slit position. In these plots pixel position 0 is to the south. {The tornadoes T1 and T2 are located between pixels 535 and 650, and 190 and 260 respectively.} Sudden drops in intensity approximately at pixels 90 and 620, seen most clearly in \ion{Mg}{2}, are due to the fiducial lines on the IRIS CCDs.}
\label{fig:MgSiC}
\end{center}
\end{figure}

The main spectral lines that are visible in the IRIS spectra are the \ion{Mg}{2} k and h lines at 2796.35 \AA\ and 2803.50 \AA\ respectively. Other lines, such as the \ion{Si}{4} 1393.78 \AA\ line and the \ion{C}{2} 1335.71 \AA\ line, appear very faint, with a low signal-to-noise ratio caused by the relatively short exposure time used in the study. We can, however, regain some information from these lines by averaging the spectral profiles along each slit over time. Figure \ref{fig:MgSiC} shows intensity as a function of slit position for the \ion{Mg}{2} k, \ion{Si}{4} 1393.71 \AA\ and \ion{C}{2} 1335.71 \AA\ lines after the profiles have been averaged for an hour of data. 

\begin{figure}
\begin{center}
\includegraphics[width=\hsize]{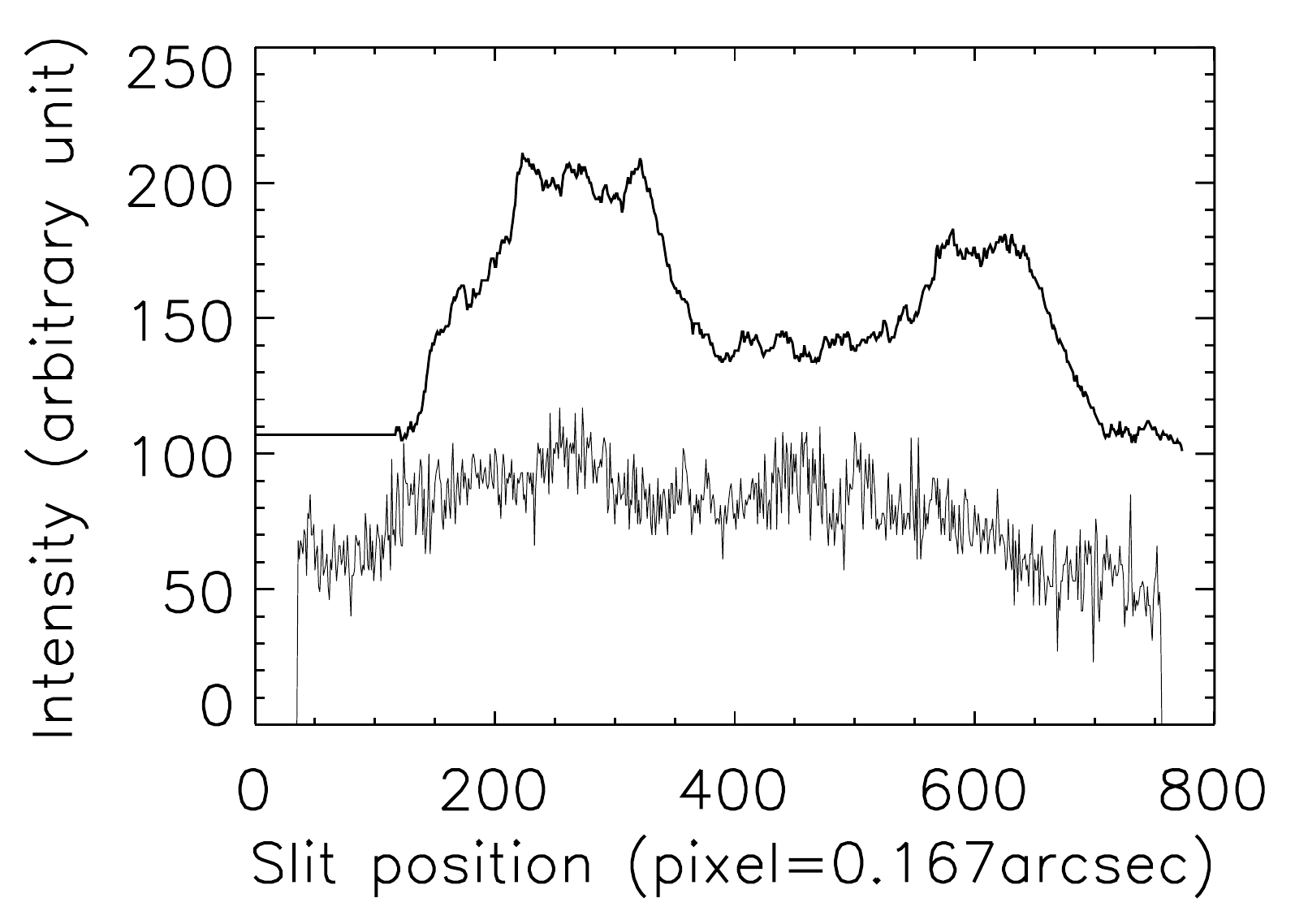}
\caption{Intensity as a function of slit position for IRIS \ion{Mg}{2} k (lower curve) and SOT \ion{Ca}{2} {H}. These are both vertical cuts through the images taken at the time of the first IRIS slit position, with the IRIS profile here coming from the slit-jaw image. {The tornadoes T1 and T2 are located between pixels 535 and 650, and 190 and 260 respectively.}}
\label{fig:MgCa}
\end{center}
\end{figure}

Performing this average we see that the \ion{Mg}{2} k line presents a very smooth profile up the slit, caused by the already strong SNR of this line. The other two lines present noisier profiles, but with an hours worth of data we begin to see patterns emerging. \ion{Si}{4} presents two bright columns, co-spatial with the dark silhouettes seen in AIA (Figure \ref{AIA_171_304}, left panel) and the brightenings seen in \textit{Hinode}/SOT \ion{Ca}{2} images (Figures \ref{fig:iris_sot} and \ref{fig:MgCa}). In the \ion{C}{2} line we see a similar profile to that of \ion{Mg}{2}.

We have analysed the \ion{Mg}{2} spectra for the raster beginning at 10:21 UT, at each of the 16 slit positions. This time was selected as it coincides with the start of the \textit{Hinode} observing time, giving us simultaneous IRIS and SOT observations. We can identify the two columns that correspond to the brightest regions in IRIS \ion{Si}{4} 1393.78 \AA\ (Figure \ref{fig:MgSiC}) and SOT \ion{Ca}{2} 3968.5 \AA\ emission (Figure \ref{fig:MgCa}), allowing us to locate them in IRIS \ion{Mg}{2} images and spectra.

\begin{figure*}
\begin{center}
\includegraphics[width=0.8\textwidth,trim=0 9cm 0 10cm,clip=true]{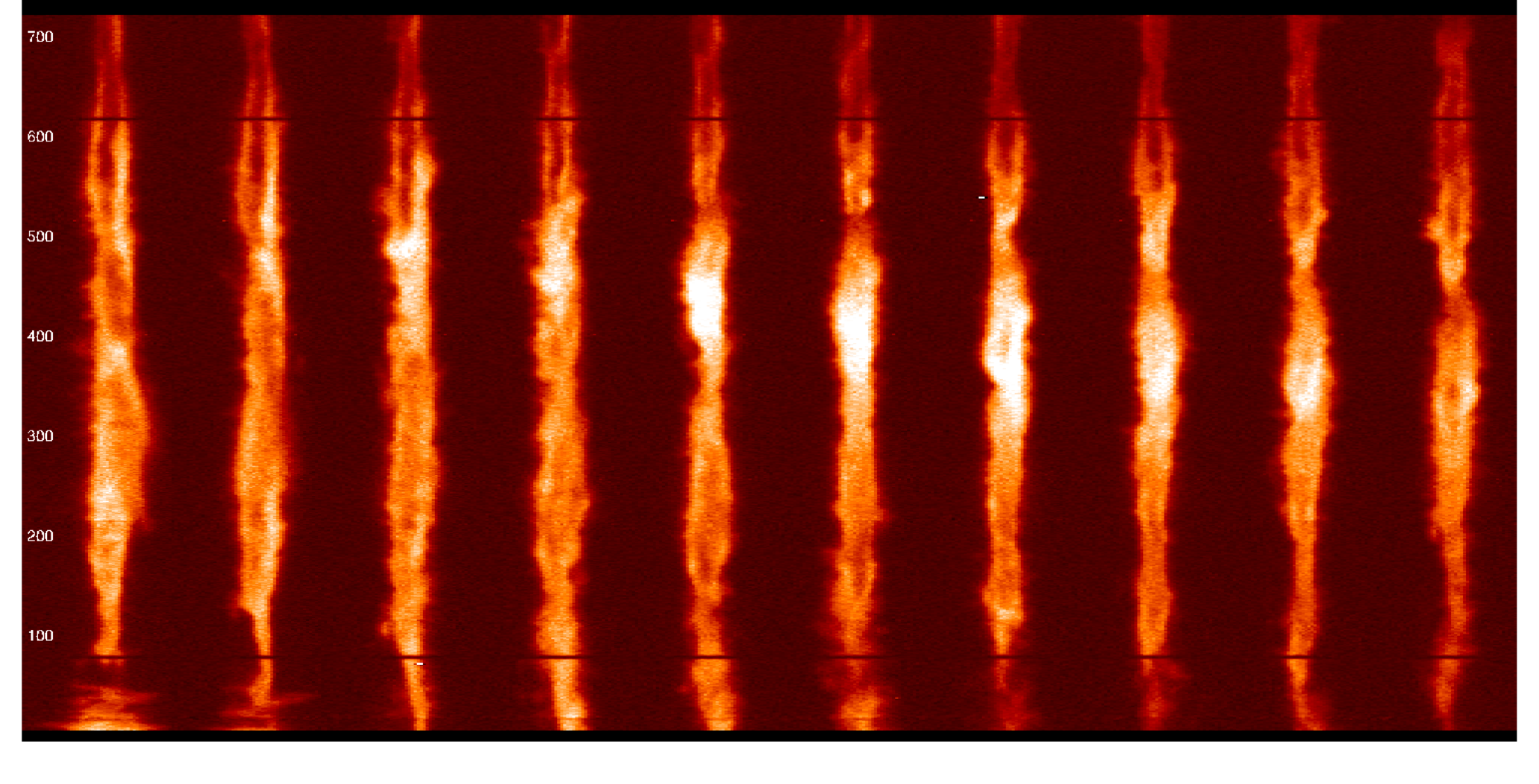}
%\caption{Examples of two Mg II h and k line spectra for three slits around 10:21 UT: slit, 1, slit 8, slit 15. The two set of arrows limit the two domain where the Ca II intensity is large in the SOT images.  The top set  corresponds to the tornado 1, the bottom set to the tornado 2. The unit  of the number on the left  is the pixel (0.167 \AA)  }
\caption{Example \ion{Mg}{2} k spectra from the raster starting at 10:21 UT. Shown here are the first 10 slit positions of the raster, with the left-most slit position being closest to the limb. The scale on the left indicates pixel number, each pixel has a size of 0.167 \AA. The horizontal dark lines through the slits approximately at pixels 90 and 620 seen here are due to the fiducial lines on the IRIS CCD. {The tornadoes T1 and T2 are located between pixels 535 and 650, and 190 and 260 respectively.}}
%\caption{Examples of the \ion{Mg}{II} h and k {spectra} for three slit {positions} in the raster starting at 10:21 UT. They are slit 1 (left), slit 8 (centre) and slit 15 (right). The two sets of arrows indicate the positions in the corresponding SOT image where the \ion{Ca}{II} intensity is the largest. The top two arrows correspond to tornado 1, the bottom two to tornado 2. The scale on the left indicates pixel number, each pixel has a size of 0.167 \AA.}
\label{fig:mg_h_k}
\end{center}
\end{figure*}

\begin{figure*}
\begin{center}
\includegraphics[width=0.9\textwidth,trim=0 7cm 0 7cm,clip=true]{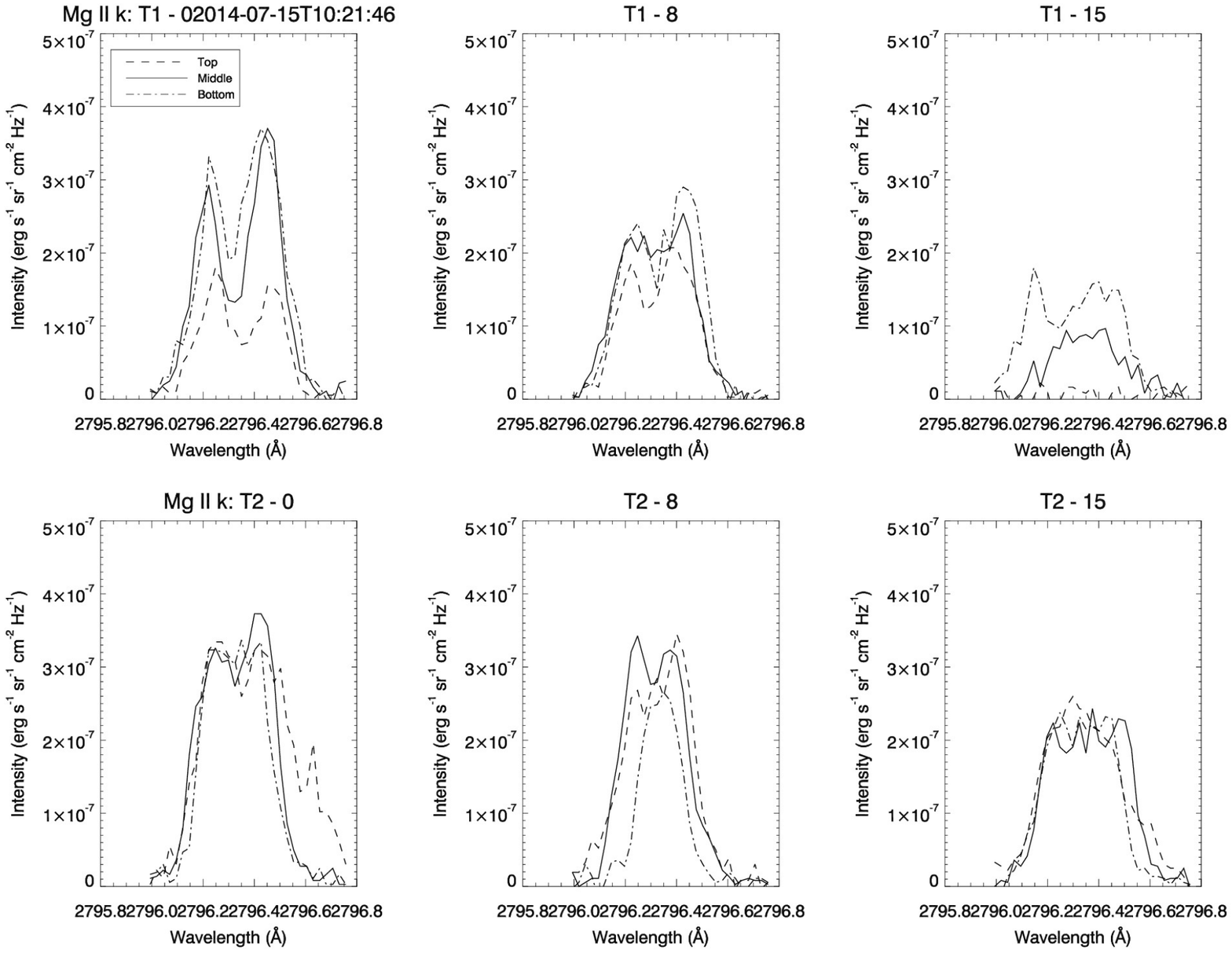}
%\caption{IRIS line profiles of Mg II 2796.4 \AA\ at 10:21 UT: {\it first row}  for pixels corresponding to  tornado 1 in slit 1, 8, 15,      {\it bottom row}  for tornado 2.}
\caption{Line profiles of the IRIS \ion{Mg}{2} k line at 10:21 UT: \textit{Top row} - spectra at three pixel locations corresponding to tornado 1, in slits 0 (left), 8 (centre) and 15 (right). \textit{Bottom row} - similar for tornado 2. Online movie shows the evolution of these profiles over time for the three hours of observation.}
\label{fig:mg_k_profiles}
\end{center}
\end{figure*}

Figure \ref{fig:mg_h_k} shows example {spectra} of the \ion{Mg}{2} k line for ten slit {positions} taken from the raster at 10:21 UT. We note that the profiles are mostly reversed or have a flat top, especially in the locations identified as being co-spatial with the tornadoes (see Figure \ref{fig:mg_k_profiles} and online movie).

\begin{figure*}
\begin{center}
\includegraphics[width=0.8\textwidth]{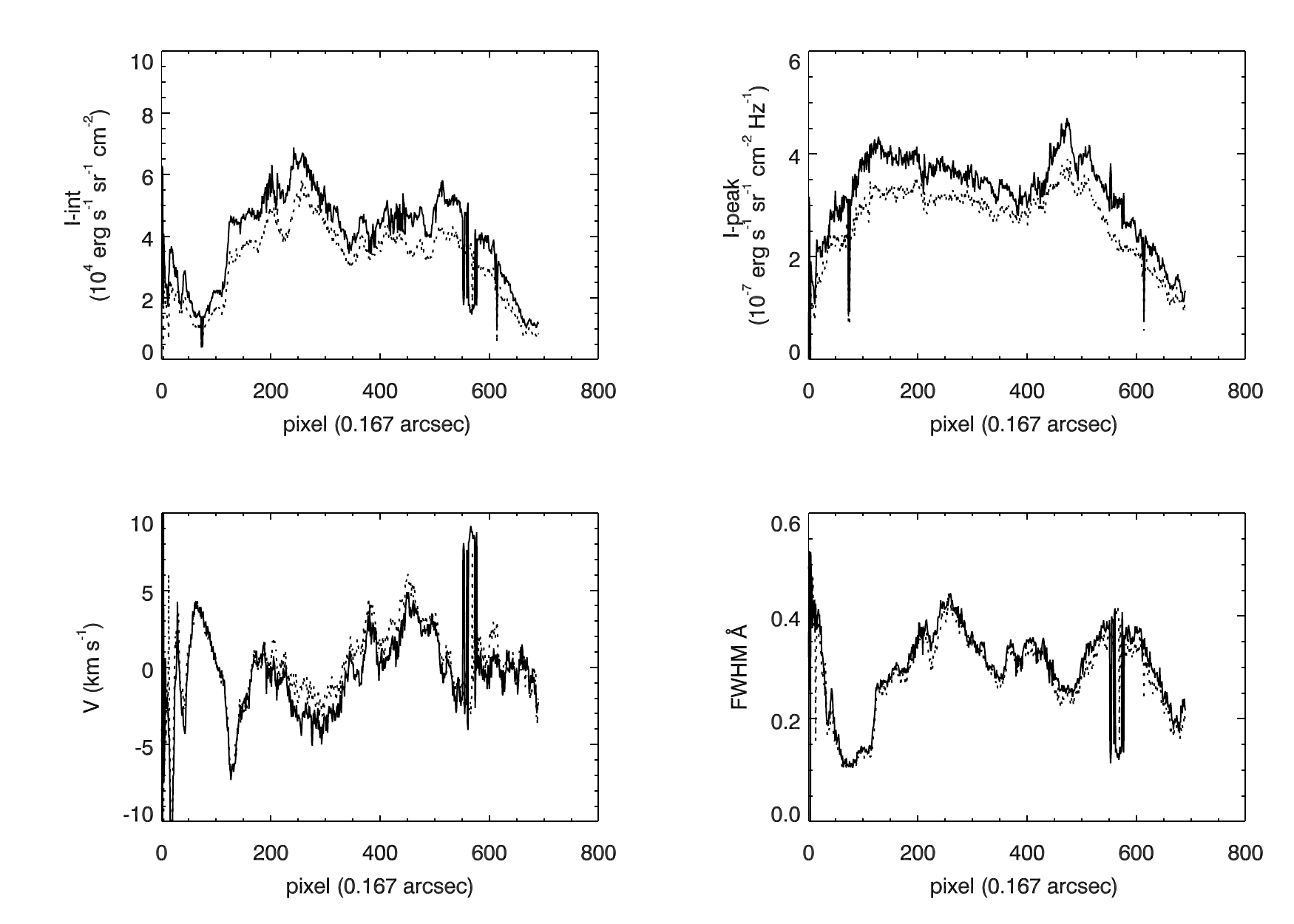}
%\caption{Charateristics of the  Mg II 2796.4 \AA\ at 10:21 UT along the slit 1:   Dopplershifts (velocity ), peak intensity, FWHM, integrated intensity assuming Gaussian profiles.              }
\caption{Fit parameters of the \ion{Mg}{2} h and k lines along slit position 1, assuming Gaussian profiles: integrated intensity, peak intensity, line-of-sight velocity and FWHM. Solid lines are for \ion{Mg}{2} k, dashed lines for \ion{Mg}{2} h. {The tornadoes T1 and T2 are located between pixels 535 and 650, and 190 and 260 respectively.}}
%\nl{Why slit 1? Also, I'd be interested to see error bars, out of curiosity. I guess we could get rid of the negative FWHM values... Need info on what solid / doted lines are. Can we have x-axis with actual arcsec position? Also show rough location of tornado(es).}}
\label{fig:mg_hk_paras}
\end{center}
\end{figure*}

Assuming a single Gaussian fit, we have computed the characteristics of the profiles for the two lines (Figure \ref{fig:mg_hk_paras}). The full widths at half maximum (FWHMs) are identical and the Doppler shifts are very similar, however the Doppler shifts between $\pm$ 5 km s$^{-1}$ are slightly higher for the h line. The FWHM values fall in a range between 0.3 \AA\ and 0.4 \AA\ with some minima ($\sim$0.18 \AA) in the top and the bottom of the slit where the profiles are not reversed. The integrated intensities are around $4 \times 10^{4}$ erg s$^{-1}$ sr$^{-1}$ cm$^{-2}$ for the k line and around $3 \times 10^{4}$ erg s$^{-1}$ sr$^{-1}$ cm$^{-2}$ for the h line. In T1, which corresponds to the top of the {slit}, the intensities decrease over a short distance to $1 \times 10^{4}$ erg s$^{-1}$ sr$^{-1}$ cm$^{-2}$. The brightest parts of the spectra correspond to the loops between the two columns. Slit 1 crosses the south column (T2) close to the limb between pixels 180 and 230, the north column near the top (T1) between pixel 570 and 630. 

Figure \ref{fig:mg_k_profiles} shows examples of the \ion{Mg}{2} profiles in T1 and T2 corresponding to slit {positions} 0, 8 and 15. The T1 profiles are reversed, {whereas} the T2 profiles show a mixture of reversed and flat-topped profiles. Table \ref{tab:line_params} shows a number of derived parameters from the line profiles of the \ion{Mg}{2} lines, derived using the analysis of the moments of the \ion{Mg}{2} distribution. The maximum of the ratio I$_{\mathrm{peak}}$/I$_0$ is found to be around 2.6 for the k line, with a corresponding integrated intensity of $4.5 \times 10^{4}$ erg s$^{-1}$ sr$^{-1}$ cm$^{-2}$.% some ratio values between the peak   center intensities. The maximum value of Ipeak/I0 is around 2.4. for a integrated intensity of  3 X10$^{4}$ erg s-1 sr-1 cm-2 Hz -1. 
\ The T2 profiles are less reversed (max = 1.4) and have higher integrated intensities of around $4.3 \times 10^{4}$ erg s$^{-1}$ sr$^{-1}$ cm$^{-2}$.

\begin{table*}
\begin{center}
\caption{Line parameters measured for the IRIS \ion{Mg}{2} k and h lines. Peak intensity has units erg s$^{-1}$ sr$^{-1}$ cm$^{-2}$ Hz$^{-1}$, and integrated intensity has units erg s$^{-1}$ sr$^{-1}$ cm$^{-2}$. FWHM is in \AA. For reversed profiles, I$_{\mathrm{peak}}$/I$_0$ is defined as the ratio of the peak intensity to the intensity at the reversal minimum. {Also shown here are values of line-of-sight velocity, v$_{\mathrm{los}}$ with units in km s$^{-1}$, as derived from a single Gaussian fit to the \ion{Mg}{2} k line.}}
\label{tab:line_params}
\begin{tabular}{ l l c c c c c c c c c }
\hline \hline
\ion{Mg}{2} k & Slit &  & 0 &  &  & 8 &  &  & 15 &   \\
\hline
 &  & North & Centre & South & North & Centre & South & North & Centre & South \\
\hline
T1 & I$_{\mathrm{peak}} ~ (\times 10^{-7})$ & 2.00 & 4.15 & 4.15 & 2.32 & 2.84 & 3.25 & 0.58 & 1.08 & 2.01 \\
 & I$_{\mathrm{peak}}$/I$_0$ & 2.41 & 2.63 & 1.89 & 1.70 & 1.95 & 1.25 & 1.72 & 1.17 & 1.62 \\
 & I$_{\mathrm{int}} ~ (\times 10^{4})$& 2.22 & 4.48 & 5.13 & 2.80 & 3.62 & 4.01 & 0.43 & 1.36 & 2.81 \\
 & FWHM & 0.31 & 0.33 & 0.31 & 0.30 & 0.29 & 0.30 & 0.08 & 0.31 & 0.35 \\
 & I$_{\mathrm{k}}$/I$_{\mathrm{h}}$ & 1.39 & 1.37 & 1.41 & 1.37 & 1.39 & 1.35 & 1.16 & 1.75 & 1.37 \\
 & v$_{\mathrm{los}}$ & -1.19 & 0.88 & 1.57 & -1.70 & -3.53 & -0.09 & -7.65 & 0.17 & -4.58 \\
\hline
T2 & I$_{\mathrm{peak}} ~ (\times 10^{-7})$ & 3.74 & 4.18 & 3.77 & 3.87 & 3.83 & 2.97 & 2.91 & 2.72 & 2.66 \\
 & I$_{\mathrm{peak}}$/I$_0$ & 1.29 & 1.36 & 1.01 (nr) & 1.21 & 1.23 & 1.00 (nr) & 1.00 (nr) & 1.33 & 1.25 \\
 & I$_{\mathrm{int}} ~ (\times 10^{4})$& 6.33 & 5.28 & 4.56 & 4.39 & 4.32 & 2.40 & 4.06 & 3.86 & 3.54 \\
 & FWHM & 0.39 & 0.30 & 0.28 & 0.30 & 0.27 & 0.25 & 0.33 & 0.33 & 0.30 \\ 
 & I$_{\mathrm{k}}$/I$_{\mathrm{h}}$ & 1.39 & 1.41 & 1.39 & 1.45 & 1.37 & 1.49 & 1.33 & 1.25 & 1.30 \\
 & v$_{\mathrm{los}}$ & 2.28 & -1.61 & -2.68 & -1.57 & -4.04 & -1.23 & -2.10 & 0.87 & -2.46 \\

\hline \hline
\ion{Mg}{2} h & Slit &  & 0 &  &  & 8 &  &  & 15 &   \\
\hline
 &  & North & Centre & South & North & Centre & South & North & Centre & South \\
\hline
T1 & I$_{\mathrm{peak}} ~ (\times 10^{-7})$ & 1.46 & 3.63 & 2.98 & 1.78 & 2.28 & 2.70 & 0.50 & 0.70 & 1.40 \\
 & I$_{\mathrm{peak}}$/I$_0$ & 2.42 & 3.21 & 1.68 & 1.40 & 1.26 & 1.64 & n/a & 1.29 & 1.63 \\
 & I$_{\mathrm{int}} ~ (\times 10^{4})$& 1.60 & 3.27 & 3.63 & 2.04 & 2.61 & 2.97 & 0.37 & 0.77 & 2.06 \\
 & FWHM & 0.35 & 0.30 & 0.30 & 0.26 & 0.28 & 0.25 & 0.41 & 0.35 & 0.33 \\
 & I$_{\mathrm{k}}$/I$_{\mathrm{h}}$ & 1.39 & 1.37 & 1.41 & 1.37 & 1.39 & 1.35 & 1.16 & 1.75 & 1.37 \\
\hline
T2 & I$_{\mathrm{peak}} ~ (\times 10^{-7})$ & 2.92 & 3.11 & 3.20 & 2.89 & 2.89 & 2.44 & 2.41 & 2.22 & 2.63 \\
 & I$_{\mathrm{peak}}$/I$_0$ & 1.07 & 1.21 & 1.12 & 1.33 & 1.42 & 1.00 (nr) & 1.19 & 1.32 & 1.57 \\
 & I$_{\mathrm{int}} ~ (\times 10^{4})$& 4.57 & 3.76 & 3.30 & 3.05 & 3.17 & 1.61 & 3.05 & 3.09 & 2.74 \\
 & FWHM & 0.34 & 0.24 & 0.23 & 0.24 & 0.28 & 0.24 & 0.30 & 0.29 & 0.27 \\ 
 & I$_{\mathrm{k}}$/I$_{\mathrm{h}}$ & 1.39 & 1.41 & 1.39 & 1.45 & 1.37 & 1.49 & 1.33 & 1.25 & 1.30 \\
\hline
\end{tabular}
\end{center}
\end{table*}

\subsubsection{Gas pressure and density}
\label{sssec:gas_pressure}
With this information we are able to discuss the gas pressure using the table of \citet{Heinzel14} for isobaric 1D models. For T1 some models could be considered with a pressure reaching 0.5 dyne cm$^{-2}$, temperatures of 6000 K and with a slab thickness of 1000 km. This would lead to an optical thickness of 820. With a lower pressure p = 0.1 dyne cm$^{-2}$ {and a temperature of 8000 K} the optical thickness could be half (460) and the geometrical depth 5000 km. This is a reasonable scenario, because we see strong absorption in the AIA 171 \AA\ images.

High optical thickness is consistent with reversal of profiles. The large absorption, or reversal, in the centre of the \ion{Mg}{2} h and k lines reduces the integrated intensity values relative to what they would be if the profiles were not reversed. This explains why T1, when viewed in the IRIS slit jaw, is not as bright as it appears in the SOT \ion{Ca}{2} images. 
For H-$\alpha$, Ca and Mg with respective optical thicknesses $\tau_{\mathrm{H-}\alpha}$, $\tau_{\mathrm{Ca}}$ and $\tau_{\mathrm{Mg}}$, when we have $\tau_{\mathrm{H-}\alpha} = 1$ we expect $\tau_{\mathrm{Ca}}$ to be 40 and $\tau_{\mathrm{Mg}}$ to be 70 (P. Heinzel, \textit{private communication}). {For \ion{He}{2} 304 \AA\ we expect $\tau_{304} \simeq$ 280 (with T = 6000 -- 8000 K, p = 0.1 dyne cm$^{-2}$, slab width = 1000 km).} This means that the line-of-sight emission in H-$\alpha$ is the sum of the emission from all the different structures along the line of sight, while in the other lines what we see is mainly what is in the frontmost part.

In T2 there is generally less reversal in the profiles, with some even appearing non-reversed. Comparing these to the table of \citet{Heinzel14} leads us to a low gas pressure scenario, with values of around 0.1 dyne cm$^{-2}$. However this is not an acceptable solution because, as mentioned previously, both the absorption in 171 \AA\ and the emission in \ion{Ca}{2} and \ion{He}{1} D$_3$ lines are strong. {Instead of the 1D, isobaric case of \citet{Heinzel14}} we have to consider 2D models with a prominence-to-corona transition region (PCTR), as was outlined in \citet{Heinzel15}. In that paper, the authors showed that these models allow us to fit H-$\alpha$ and \ion{Mg}{2} lines with a relatively high pressure. Some of the profiles could also be interpreted by fitting with multiple Gaussians. This would indicate that the line of sight is crossing several structures, each with a different velocity - as was shown in \citet{Schmieder14}.

\subsubsection{Doppler shifts}
\label{sssec:doppler_shifts}
%The Doppler shifts using the assumption of gaussian line profiles lead to small velocities (+/-5 km/s).  The pattern obtained for the velocities follow the intensity pattern of the loops between the two columns with alternatively red and blue shifts (Figure maps of dopplershifts).  The evolution versus time is slow . In one hour we see the blueshift on one side of the loop has reached the other  end of the loop.   Working on the moments of the line profiles, the pattern is unchanged. The values are of the same order. 

\begin{figure*}
\begin{center}
%\includegraphics[width=0.49\textwidth]{iris_vel_08:30.eps}
%\includegraphics[width=0.49\textwidth]{iris_vel_09:04.eps}
%\\
%\includegraphics[width=0.49\textwidth]{iris_vel_10:16.eps}
%\includegraphics[width=0.49\textwidth]{iris_vel_10:34.eps}
%\includegraphics[width=0.49\textwidth]{IRIS_20140715_08:30_new.eps}
%\includegraphics[width=0.49\textwidth]{IRIS_20140715_09:04_new.eps}
%\\
%\includegraphics[width=0.49\textwidth]{IRIS_20140715_10:16_new.eps}
%\includegraphics[width=0.49\textwidth]{IRIS_20140715_10:34_new.eps}
\includegraphics[width=0.21\textwidth,trim=5cm 0 6cm 0,clip=true]{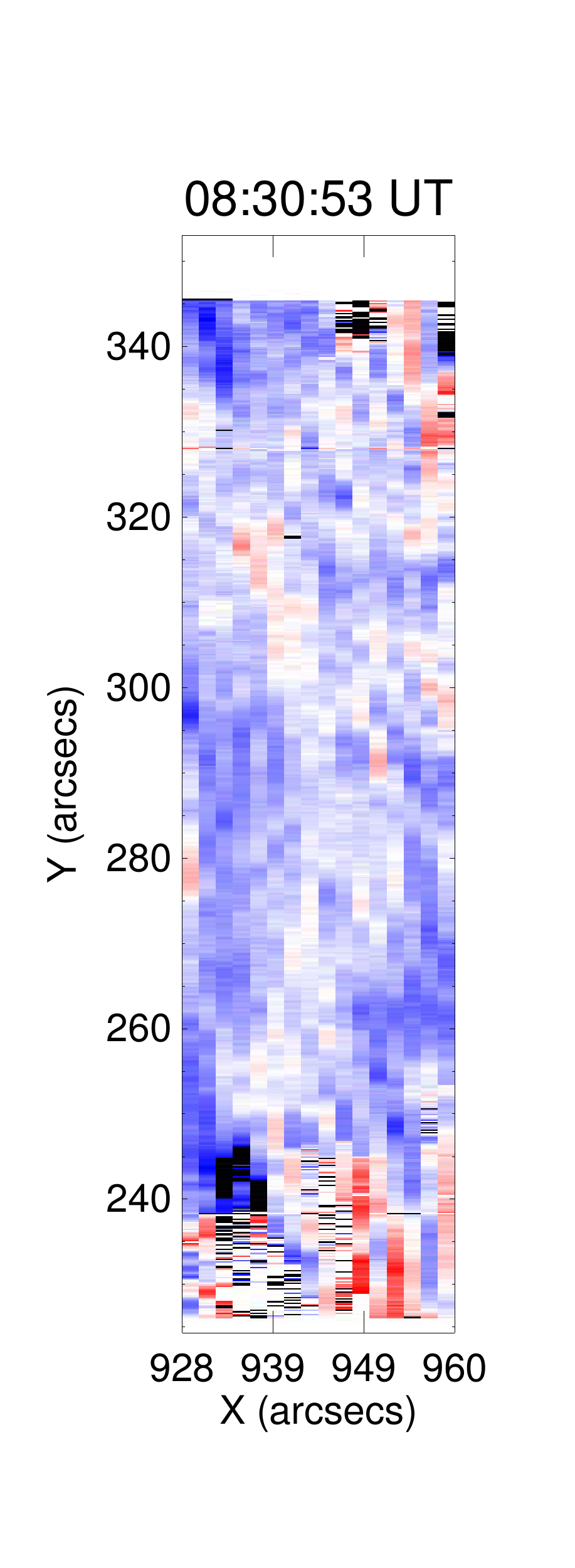}
\includegraphics[width=0.21\textwidth,trim=5cm 0 6cm 0,clip=true]{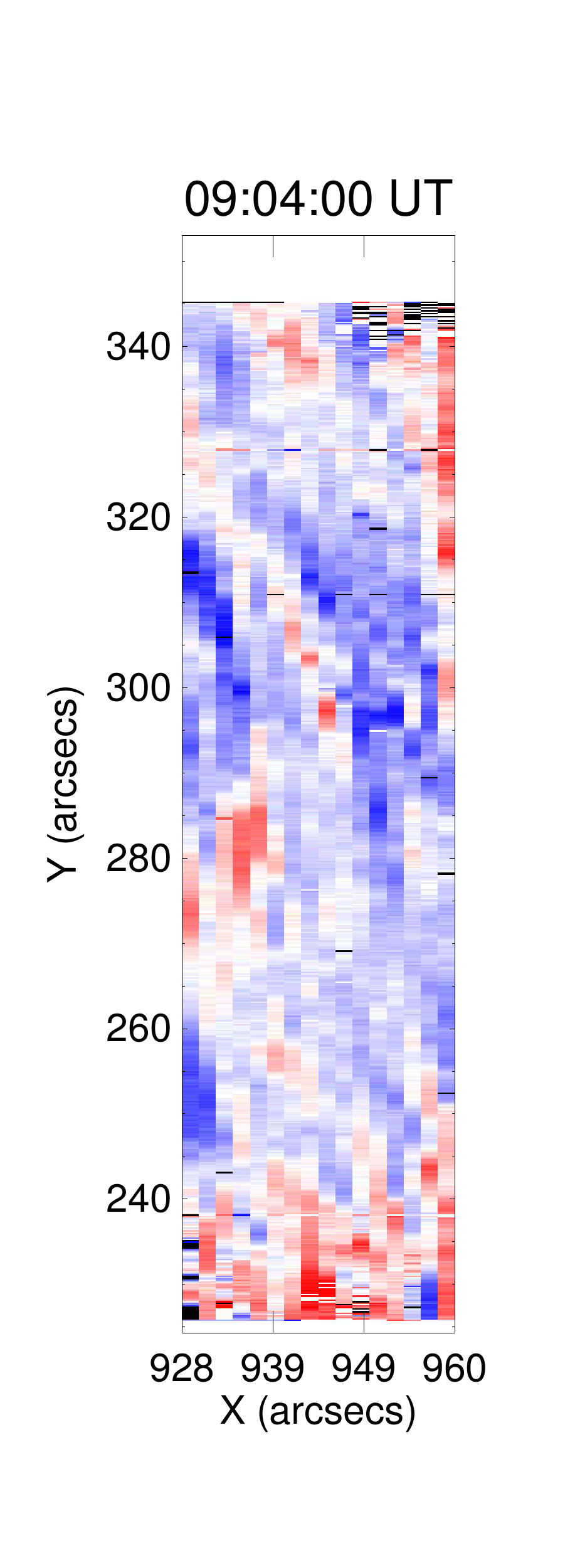}
\includegraphics[width=0.21\textwidth,trim=5cm 0 6cm 0,clip=true]{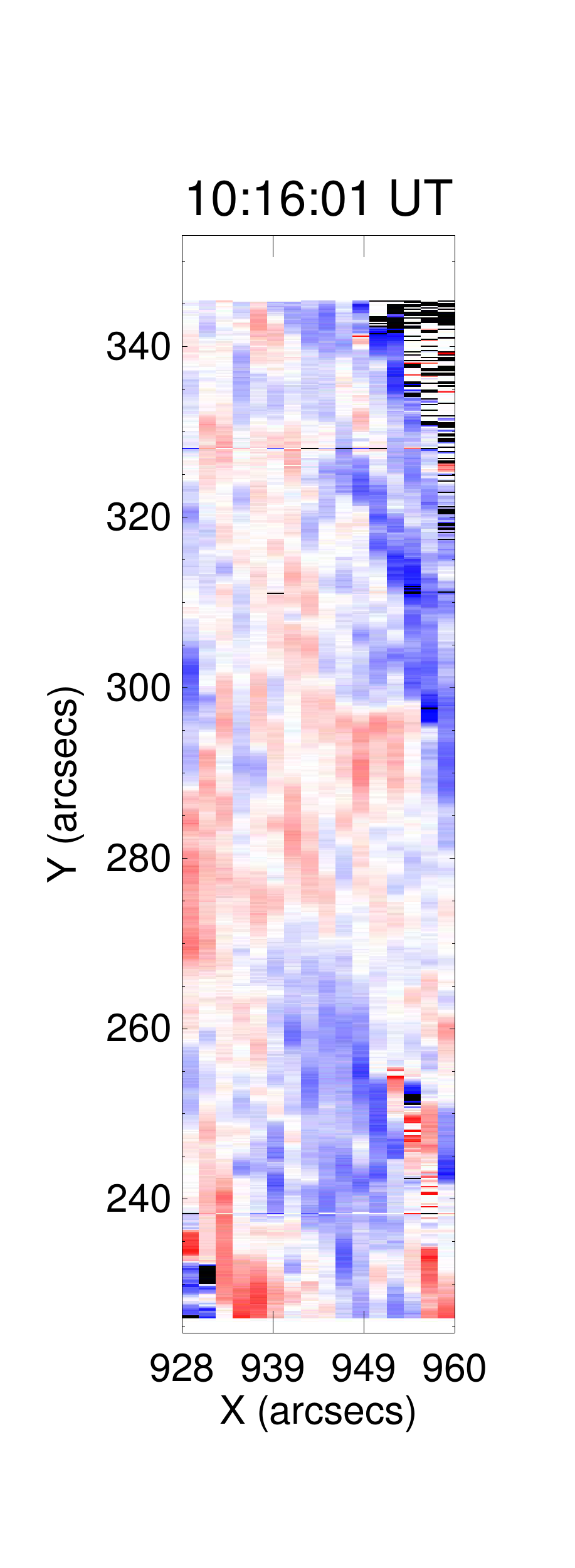}
\includegraphics[width=0.21\textwidth,trim=5cm 0 6cm 0,clip=true]{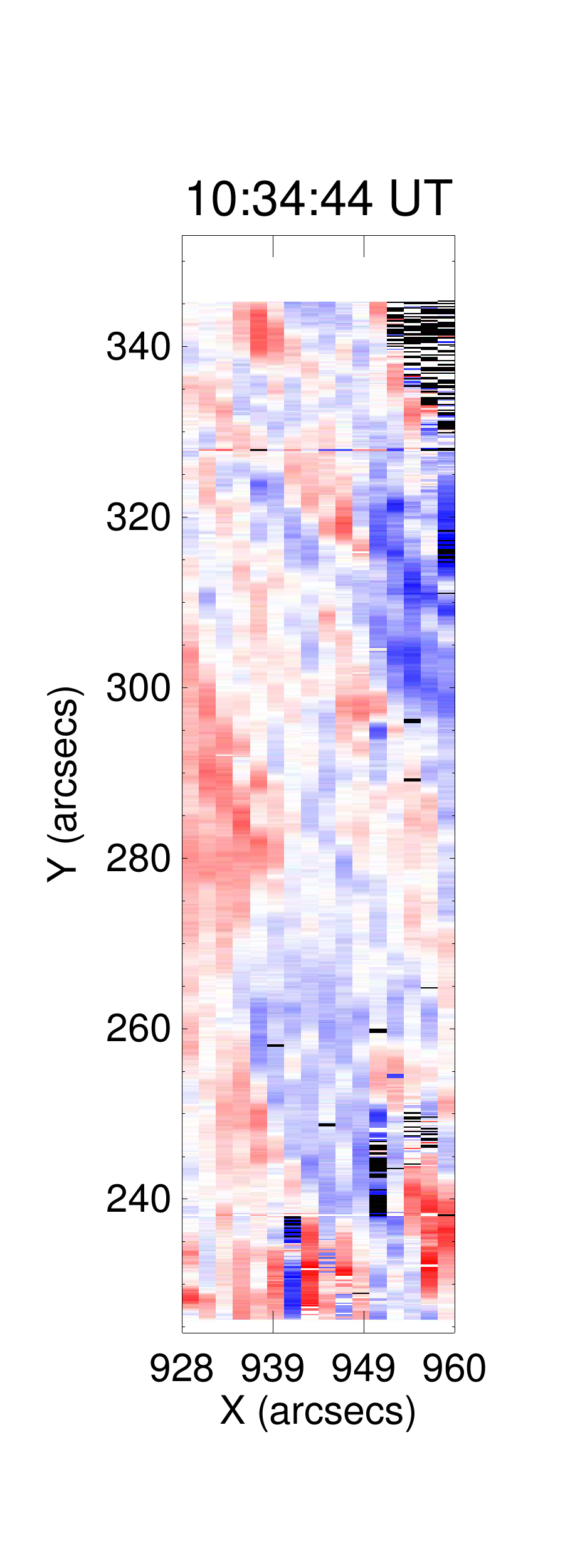}
\caption{IRIS line-of-sight velocity maps at four different times during the study, made using the Gaussian assumption for the \ion{Mg}{2} k line. Limits for the velocity in these plots are $\pm~10$ km s$^{-1}$.}
\label{fig:IRISvels}
\end{center}
\end{figure*}

\begin{figure*}
\begin{center}
\includegraphics[width=0.45\textwidth]{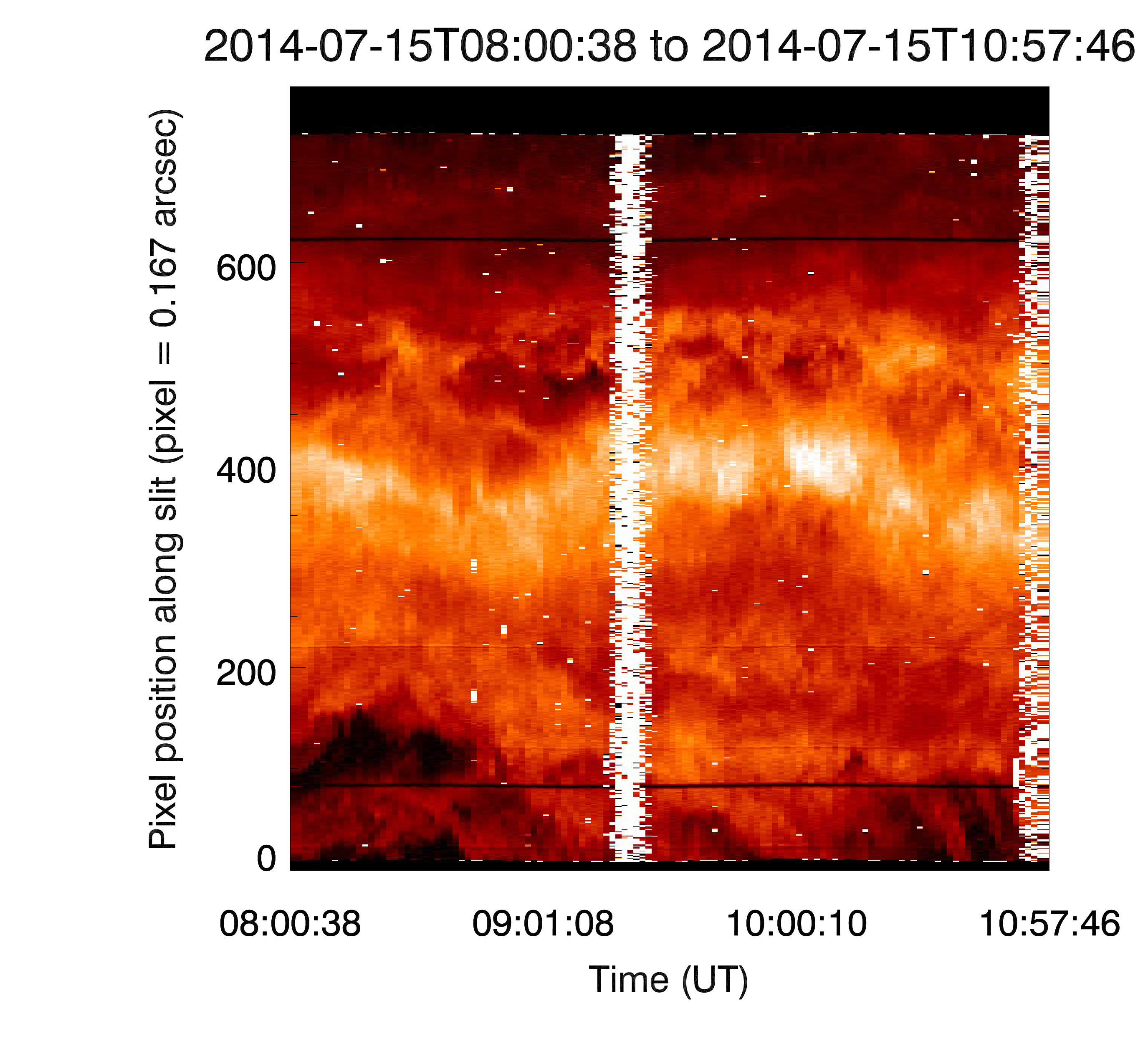}
\includegraphics[width=0.45\textwidth]{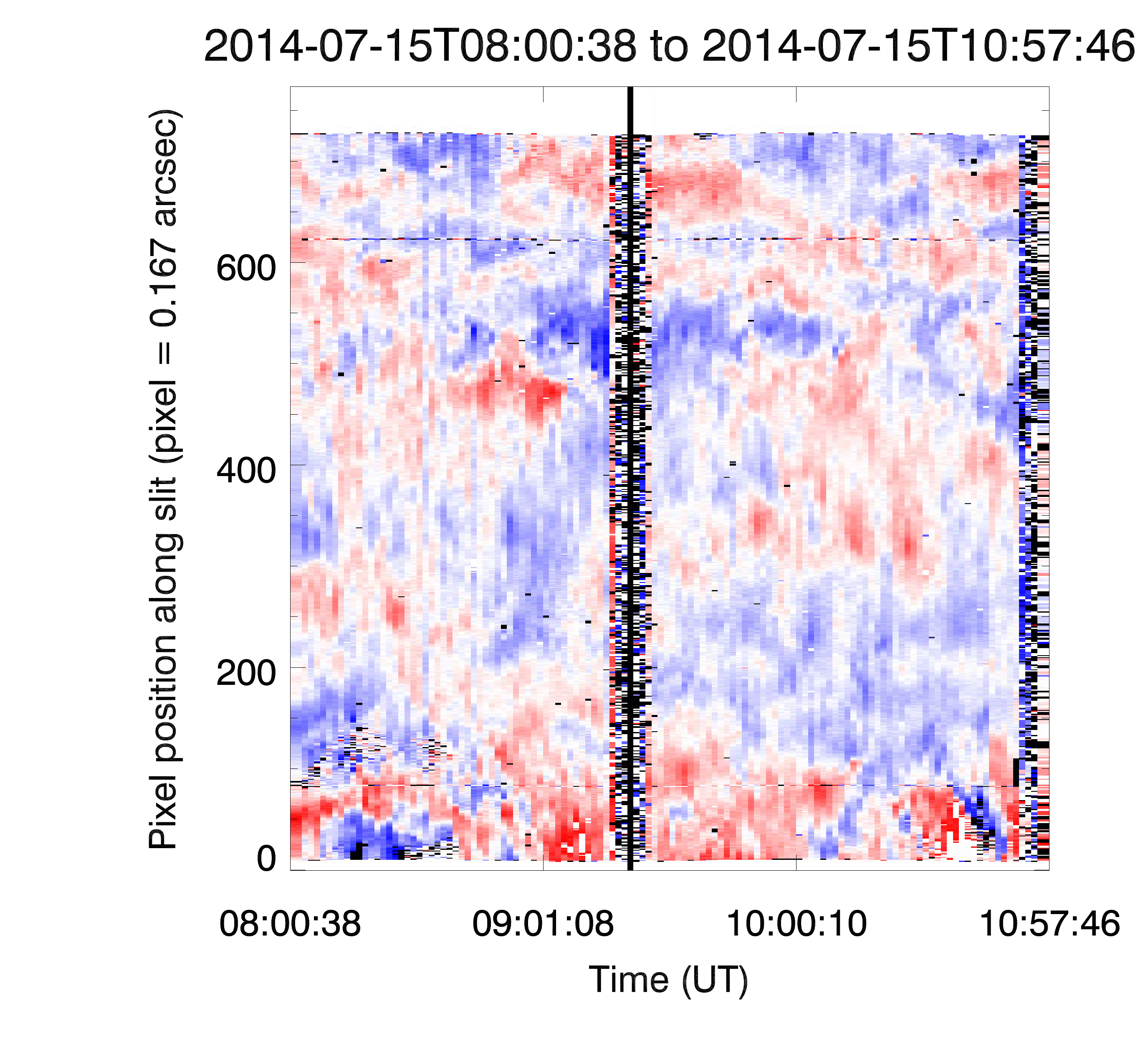}
\caption{Time evolution of the peak intensity (left) and velocity (right) along the central slit of the IRIS raster. The tornadoes T1 and T2 are located between pixels 535 and 650, and 190 and 260 respectively. The vertical noise dominated regions are caused by the spacecraft passing over the South Atlantic Anomaly.}
\label{fig:IRIS_time_distance}
\end{center}
\end{figure*}

Assuming Gaussian profiles for the IRIS \ion{Mg}{2} lines, we return relatively small line-of-sight velocities ($\pm$ 5 km s$^{-1}$). The velocity pattern obtained using this method appears to follow the intensity pattern of the loops inbetween the two columns, seen in Figure \ref{fig:IRISvels}, which alternates between red and blueshifts when viewed over time (see Figure \ref{fig:IRIS_time_distance}). This evolves slowly over time, taking about an hour for the blueshifted part to move from one end of the loop to the other. If we instead look at an analysis of the moments of the distributions we find similar intensity and velocity patterns, with values of the same order as those found by Gaussian fitting.

{The bottom left-hand panel of Figure \ref{fig:mg_hk_paras} shows the velocity along one slit using a single Gaussian fit for both the \ion{Mg}{2} k and h lines. The middle of T2 is around pixel 200, with the centre of T1 being around pixel 600. The bottom of the slit corresponds to pixel 0. There is a velocity gradient around T2, but this is not enough evidence in support of rotation. Around T1 we see large spikes of velocity, which is where the Gaussian fitting fails for the most reversed profiles. }

Looking at the spectral images and individual profiles, we notice some undulation in both the blue and red wings along one slit (Figure \ref{fig:mg_k_profiles} and profiles from the top of the spectra). These undulations are due to variations in the relative blue and red wing intensities, leading to Doppler shifts on the order of 10 km s$^{-1}$. Some profiles are also shifted by up to 20 km s$^{-1}$ (Figure \ref{fig:mg_k_profiles}, T1 - 15), however these are transient and do not remain for longer than one or two rasters.

\section{Discussion}
\label{sec:disc}

{Solar tornadoes and tornado-like prominences have been the topic of a number of publications recently, especially since the launch of the \textit{SDO} spacecraft with its high-resolution imager AIA. Spectrometers such as those on the \textit{Hinode} and IRIS spacecraft have furthered our understanding of these features, but they have also raised many more questions themselves.

We have here shown results from a coordinated observation of a solar tornado from 15 July 2014. We see two tornado-like features on the limb, and we have presented analysis of this event using data from a number of instruments - \textit{SDO}/AIA, \textit{Hinode}/SOT, IRIS, and THEMIS.

Time-distance analysis of our event using \textit{SDO}/AIA shows periods of around 1 -- 1.5 hours (Section \ref{sssec:time_distance} and Figure \ref{fig:AIA_tornado_cuts}). If the structure is rotating, this would give velocities of around 10 km s$^{-1}$, a similar order as that found by \citet{Su12}. However, \citet{Panasenco14} argue that such apparent rotation could be in fact oscillation projected onto the plane of the sky, an explaination that we cannot rule out with this data set.}

We have presented magnetic field inversions from THEMIS using the \ion{He}{1} D$_3$ lines. From these inversions we can gain information about the magnetic field strength and orientation inside the prominence structure. We find the strongest magnetic fields (up to 50 G) in the prominence legs. We also find that the magnetic field is mostly horizontal (parallel to the limb) in these tornado-like columns.

These observations do not fit with the vertical magnetic field structure that has been suggested by previous authors, pointing more towards the horizontal field structure which was explored by \citet{Dudik12}. 
\citet{Luna15} have recently suggested a model of a twisted, or `tornado-like', magnetic field structure with a small slope to the spiral. Field lines close to the axis are mostly vertical, while the field lines close to the edge of the structure are helicoidal. From the inversion of the Stokes profiles we find a nearly constant LOS azimuth of 60$^\circ$ in T2, while it is varying in T1 (Fig.~\ref{themis2}). In both cases, the inversion is affected by the 180 degree ambiguity. Hence we can confirm that we observe a horizontal magnetic field, but we do not have supporting evidence concerning the presence of the mixed azimuth distribution that is suggested by \citet{Luna15}.

{Analysis has been done on data from the 15 July 2014 using the IRIS spacecraft. We mostly present analysis from the \ion{Mg}{2} k and h lines, although we also make use of the \ion{Si}{4} and \ion{C}{2} lines. We find that the bright prominence legs that are visible in H-$\alpha$ (from ground-based observatories), IRIS \ion{Si}{4} and \ion{Ca}{2} (from \textit{Hinode}/SOT) are not readily visible in the \ion{Mg}{2} lines.}

We have presented line parameters and intensity ratios from analysis of moments of the \ion{Mg}{2} profiles in Table \ref{tab:line_params}. FWHM values are consistent with predictions from radiative transfer models such as those of \citet{Paletou93} and \citet{Heinzel14}, as well as previous observations by \citet{Vial82}. Our findings show broader profiles than were found by \citet{Schmieder14} using IRIS, however we note that in that paper the \ion{Mg}{2} profiles were not reversed and they did not observe the same dark features in AIA images.

Also shown in Table \ref{tab:line_params} are intensity ratios of peak intensity to intensity at central reversal (I$_{\mathrm{peak}}$/I$_0$) for the \ion{Mg}{2} k line, and k to h intensity ratio (I$_{\mathrm{k}}$/I$_{\mathrm{h}}$). The value given by I$_{\mathrm{peak}}$/I$_0$ tells us about the level of reversal in the profile. Here we find that the profiles in T1 are generally more reversed than in T2. The k to h intensity ratio is as expected -- models such as \citet{Paletou93} quote a k/h ratio of around 1.35, and from \citet{Heinzel14} it is expected to be around 1.37 -- 1.44. Our values also match well with previous observations, such as those in \citet{Vial82} and \citet{Schmieder14}.

The level of reversal of the profile allows us to compare with models in order to constrain the gas pressure and optical thickness of the prominence. Comparing our results to a 1D isothermal-isobaric radiative transfer model \citep[][table 2]{Heinzel14}, for a prominence slab of geometrical thickness of 1000 km and temperature 6000 K, we get a gas pressure, $\rho$, of 0.1 -- 0.5 dyne cm$^{-2}$ and an optical thickness, $\tau_{\mathrm{Mg}}$, of 140 -- 820. However, for T2 (corresponding to the lower limit of $\rho$ and $\tau_{\mathrm{Mg}}$) we note that the absorption in AIA 171 \AA\ is strong, as is the emission in \ion{Ca}{2} and \ion{He}{1} D$_3$. We therefore must consider 2D models which contain a prominence-to-corona transition region.

{We have analysed line-of-sight velocities in the \ion{Mg}{2} line (formed at chromospheric temperatures) from IRIS, to study Doppler velocities at lower plasma temperatures. Previously, \citet{Su14} and \citet{Levens15} presented results showing a split Doppler pattern in a tornado-like prominence at coronal temperatures ($>$ 1 MK). \citet{Orozco12} and \citet{Wedemeyer13} showed a similar Doppler pattern in the \ion{He}{1} 10830 \AA\ and H-$\alpha$ lines respectively, but little work has been done on lines formed at temperatures of $\log{\mathrm{T}} = 4.5$. These results are consistent with what has been obtained in other studies comparing optically thick lines with various radiative transfer models \citep{Labrosse10}.

In the Doppler maps of \ion{Mg}{2} k we find no such split pattern. We see redshifts and blueshifts on the order of $\pm$ 10 km s$^{-1}$, but these seem to be associated with the loop-like structure that is seen in intensity maps of this ion (Figure \ref{fig:IRISvels}). There is also no consistant split pattern when we look at the evolution over time (Figure \ref{fig:IRIS_time_distance}). These results do not support any of the rotating `tornado' models, however we must acknowledge that the \ion{Mg}{2} k and h lines are optically thick (as discussed in Section \ref{sssec:gas_pressure}). This means that the radiation we receive is only from the frontmost layers of the prominence, as opposed to integrated all the way along the line of sight through the tornado structure as in the optically thin case.}

\section{Conclusions}
\label{sec:conc}

We have presented results from a coordinated prominence observation on 15 July 2014 using the \textit{SDO}/AIA, IRIS and \textit{Hinode}/SOT instruments as well as the THEMIS spectropolarimeter. The prominence studied here was chosen due to its tornado-like nature, and good coordination between the instruments covering two tornado-like prominence legs.

The prominence legs appear as dark, silhouetted features in AIA coronal filters (171 \AA, 193 \AA\ etc.) due to the absorption of background emission by material contained in the prominence. These columns appear bright in IRIS \ion{Si}{4}, \textit{Hinode}/SOT \ion{Ca}{2} and THEMIS \ion{He}{1} D$_3$ lines, but we see no such brightenings in IRIS \ion{Mg}{2} line profiles. We do, however, find central reversal at these positions in \ion{Mg}{2} which is not present elsewhere in the raster.

{Comparing \ion{Mg}{2} profiles observed by IRIS to those expected from radiative transfer models, both 1D \citep{Heinzel14} and 2D \citep{Paletou93}, we find a good agreement with our data. These models allow us to constrain physical parameters of the plasma that we observe. We recover gas pressures of around 0.1 -- 0.5 dyne cm$^{-2}$ and high optical thicknesses of between 140 and 820 for the \ion{Mg}{2} k and h lines. Qualitatively, the best agreement is obtained with 2D models which contain a prominence-to-corona transition region.}

We performed line-of-sight velocity analysis of the \ion{Mg}{2} k line and found no evidence for rotation in this line. {Time-distance analysis of AIA images reveal that there are oscillations in the two tornadoes over a period of around 1 -- 1.5 hours. It is not possible to conclude what mechanism causes these oscillations from these observations.

THEMIS provides magnetic field information in these structures, both orientation and field strength. We recover field strengths that are strongest in the two tornado columns, on the order of 20 -- 50 G. The inclination of the field tells us that the field is horizontal everywhere, parallel to the limb. The line-of-sight azimuth reveals values of between 70$^\circ$ and 100$^\circ$.}

{Coordinated observation between THEMIS and other instruments provide us with a unique opportunity to study the magnetic field and plasma properties in prominence legs. We plan to extend our investigation of these dynamic structures in a future paper.}

\begin{acknowledgements}

The authors thank S. Gunar, B. Gelly and the team of THEMIS for assisting with the observations. {The authors also thank the anonymous referee for their comments which helped improve the clarity of this paper.} P.J.L. acknowledges support from an STFC Research Studentship ST/K502005/1. N.L. acknowledges support from STFC grant ST/L000741/1. \textit{Hinode} is a Japanese mission developed and launched by ISAS/JAXA, with NAOJ as domestic partner and NASA and STFC (UK) as international partners. It is operated by these agencies in co-operation with ESA and NSC (Norway). IRIS is a NASA small explorer mission developed and operated by LMSAL with mission operations executed at NASA Ames Research center and major contributions to downlink communications funded by the Norwegian Space Center (NSC, Norway) through an ESA PRODEX contract. The AIA data are provided courtesy of NASA/\textit{SDO} and the AIA science team.

\end{acknowledgements}

%-------------------------------------------------------------------

\bibliographystyle{aa}
\bibliography{bibliography}

\end{document}